\documentclass[american,english,fleqn,10pt]{wlscirep}
\pdfoutput=1
\usepackage[utf8]{inputenc}
\usepackage[T1]{fontenc}
\usepackage{gensymb}
\usepackage{multirow}
\usepackage{array}
\usepackage{amsmath}
\usepackage[caption=false]{subfig}
\usepackage{color}
\usepackage{babel}
\usepackage{float}
\usepackage{units}
\title{Excited States and Optical Properties of Hydrogen-Passivated Rectangular Graphenes: A Computational Study}

\author[1]{Deepak Kumar Rai}
\author[1,*]{Alok Shukla}
\affil[1]{Department of Physics, Indian Institute of Technology Bombay, Powai, Mumbai 400076, India}

\affil[*]{shukla@phy.iitb.ac.in}

\begin{abstract}
{\normalsize{}In this paper, we perform large-scale electron-correlated
calculations of optoelectronic properties of rectangular graphene-like
polycyclic aromatic hydrocarbon molecules. Theoretical methodology
employed in this work is based upon Pariser-Parr-Pople (PPP) $\pi$-electron
model Hamiltonian, which includes long-range electron-electron interactions.
Electron-correlation effects were incorporated using multi-reference
singles-doubles configuration-interaction (MRSDCI) method, and the
ground and excited state wave functions thus obtained were employed
to calculate the linear optical absorption spectra of these molecules,
within the electric-dipole approximation. As far as the ground state
wave functions of these molecules are concerned, we find that with
the increasing size, they develop a strong diradical open-shell character.
Our results on optical absorption spectra are in very good agreement
with the available experimental results, outlining the importance
of electron-correlation effects in accurate description of the excited
states. In addition to the optical gap, spin gap of each molecule
was also computed using the same methodology. Calculated spin gaps
exhibit a decreasing trend with the increasing sizes of the molecules,
suggesting that the infinite graphene has a vanishing spin gap. }{\normalsize\par}
\end{abstract}
\begin{document}

\flushbottom
\maketitle
%
%
\thispagestyle{empty}


\section*{{\normalsize{}Introduction}}

Despite many attractive properties, graphene has still not found applications
in opto-electronic devices, because of the lack of a band gap. Therefore,
in recent times, considerable amount of research effort has been directed
towards graphene nanostructures such as quantum dots,\cite{quantum-dot-exp-review,gqd-exp-review-choi,sk2014revealing,yamijala2015linear,geethalakshmi2016tunable,plasser2013multiradical}
and nanoribbons,\cite{nanoribbon-exp-review} which are expected
to have band gaps because of quantum confinement. The idealized graphene
nanoribbons (GNRs) have either zigzag or armchair edges, with substantially
different electronic structure, and related properties. Theoretical
studies reveal that zigzag GNRs (ZGNRs) exhibit edge magnetism with
possible applications in spintronic devices,\cite{zgnr-nature-theory-louie,zgnr-prl-theory-louie}
while the armchair GNRs (AGNRs) are direct bandgap semiconductors,
with potential optoelectronic applications.\cite{agnr-theory-scuseria,agnr-theory-louie}
If we consider either an AGNR or a ZGNR of a given width, and hypothetically
cut it at two places perpendicular to its width the resultant rectangular
structure, referred to as a rectangular graphene molecule (RGM), will
have both armchair and zigzag type edges. We also assume that the
edge carbon atoms of RGMs are passivated by H atoms, so as to neutralize
the dangling bonds, thus preventing edge reconstruction, and allowing
them to retain their symmetric shapes. Such structures, obviously,
will be nothing but polycyclic aromatic hydrocarbon molecules. In
this work we perform a computational study of optoelectronic properties
of RGMs, with the aim of understanding as to how they are influenced
by the edge structure. Because the electronic properties of ZGNRs
and AGNRs are very different from each other, it is of considerable
interest as to how the electronic properties of RGMs, which have both
zigzag and armchair edges, evolve with the edge lengths. Such an understanding
will help us in tuning the optoelectronic properties of RGMs by manipulating
their edges.

In our theoretical approach we consider RGMs to be systems whose low-lying
excited states are determined exclusively by their $\pi$ electrons,
with negligible influence of $\sigma$ electrons. As a result we adopt
a computational approach employing the Pariser-Parr-Pople (PPP) $\pi-$electron
Hamiltonian,\cite{ppp-pople,ppp-pariser-parr} and the configuration
interaction (CI) method, used in several of our earlier works on conjugated
polymers,\cite{PhysRevB.65.125204Shukla65,PhysRevB.69.165218Shukla69,PhysRevB.71.165204Priya_t0,:/content/aip/journal/jcp/131/1/10.1063/1.3159670Priyaanthracene,doi:10.1021/jp408535u,himanshu-triplet,sony-acene-lo}
polycyclic aromatic hydrocarbons,\cite{doi:10.1021/jp410793rAryanpour,:/content/aip/journal/jcp/140/10/10.1063/1.4867363Aryanpour}
and graphene quantum dots.\cite{springerchapter,Tista1,Tista2}
We adopt this approach to study RGMs with the number of carbon atoms
ranging from 28 to 56, corresponding to structures with increasing
edge lengths in both armchair, and zigzag, directions. Adopting the
notation that RGM-$n$ denotes a rectangular graphene structure with
$n$ carbon atoms, the chemical analogs of RGM-28, -30, -36, -40,
-42, -50, and -56 are aromatic compounds bisanthenes, terrylene, tetrabenzocoronene,
quaterrylene, teranthene, pentarylene and quateranthene, respectively.
On comparing our theoretical results to the measured ones on these
molecules, we obtain excellent agreement, thus validating our methodology.

Additionally, using the same MRSDCI methodology, we computed the spin
gap of each RGM studied in this work. We find that with the increasing
sizes of the RGMs, their spin gaps are decreasing, suggesting that
the spin gap of infinite graphene vanishes.

\subsection*{Structure and Symmetry}

\label{sec:symmetry}

\begin{figure}
\begin{centering}
\includegraphics[scale=0.4]{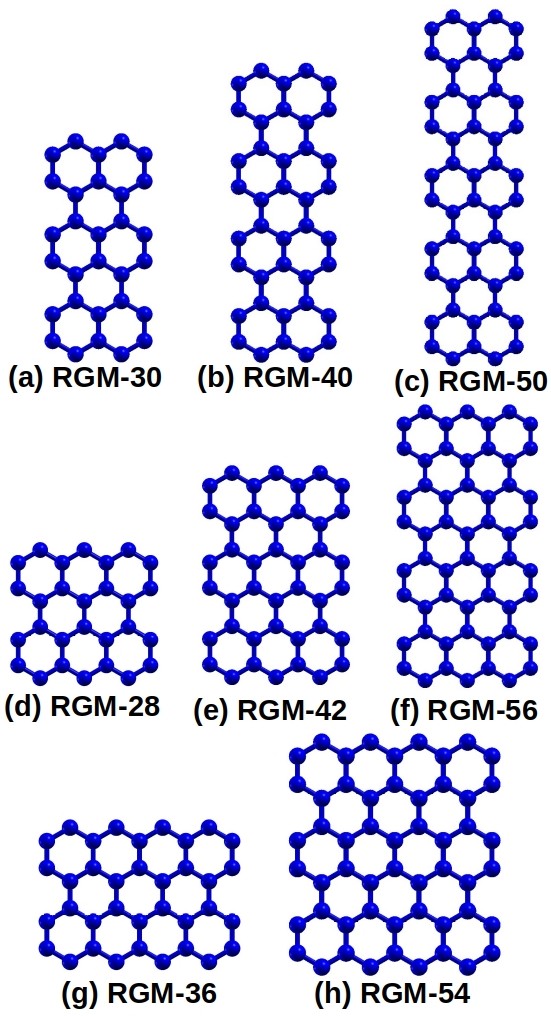}
\par\end{centering}
\caption{\label{fig:Schematic-diagram-RGQDs_w_varying_length}Schematic diagrams
of RGMs considered in this work. For all the molecules, edge carbon
atoms are assumed passivated by hydrogens. Notation RGM-$n$ denotes
a rectangular-shaped graphene-like molecule with $n$ carbon atoms.}
\end{figure}

The schematic diagrams of RGMs considered in this work are shown in
Fig. \ref{fig:Schematic-diagram-RGQDs_w_varying_length}. As mentioned
earlier, we have assumed that the dangling bonds on the edges of the
molecules have been saturated by hydrogen atoms. Thus, the molecules
considered here can be treated as planar hydrocarbons, exhibiting
$\pi$ conjugation. We have assumed that all the RGMs lie in the $xy$
plane, with idealized bond lengths of 1.40 \AA, and bond angles of
$120^{{\degree}}$. 

Within the PPP model based theoretical methodology adopted here, small
variations in bond lengths and angles do not make any significant
differences to the calculated optical properties of such structures,
as demonstrated by us earlier.\cite{doi:10.1021/jp408535u,himanshu-triplet}
Having assumed the idealized geometry for the RGMs, their point group
symmetry is $D_{2h}$, as in case of polyacenes studied by us earlier.\cite{sony-acene-lo,:/content/aip/journal/jcp/131/1/10.1063/1.3159670Priyaanthracene,doi:10.1021/jp408535u,himanshu-triplet}
Because all the systems considered here have an even number of electrons,
their ground state is of $^{1}A_{g}$ symmetry, so that their one-photon
dipole-connected excited states will be of the symmetries: (a) $^{1}B_{3u}$,
accessible by absorbing an $x$-polarized photon, and (b) $^{1}B_{2u}$,
reached through the absorption of a $y$-polarized photon. 

\section*{{\normalsize{}Theoretical Methods}}

\label{sec:theory}

As discussed in the previous section, with hydrogen passivated edges,
the molecules considered here are $\pi$-conjugated systems, and,
therefore, in this work we employed Pariser-Parr-Pople (PPP) effective
$\pi$-electron model Hamiltonian,\cite{ppp-pople,ppp-pariser-parr} 

\begin{align}
H & \mbox{\ensuremath{=}\ensuremath{-}\ensuremath{\sum_{i,j,\sigma}t_{ij}\left(c_{i\sigma}^{\dagger}c_{j\sigma}+c_{j\sigma}^{\dagger}c_{i\sigma}\right)}\ensuremath{+}\ensuremath{U\sum_{i}n_{i\uparrow}n_{i\downarrow}}}\nonumber \\
\mbox{\mbox{\mbox{}}} & \mbox{+\ensuremath{\sum_{i<j}V_{ij}(n_{i}-1)(n_{j}-1),}}\label{eq:ppp}
\end{align}

where $c_{i\sigma}^{\dagger}($c$_{i\sigma})$ are creation (annihilation)
operators for the $p_{z}$ orbital of spin $\sigma$, located on the
$i$-th carbon atom, while the total number of electrons on the atom
is indicated by the number operator $n$$_{i}=\sum_{\sigma}c_{i\sigma}^{\dagger}c_{i\sigma}$.
The first term in Eq. \ref{eq:ppp} denotes the one-electron hopping
processes connecting $i$-th and $j$-th atoms, quantified by matrix
elements $t_{ij}$. We assume that the hopping connects only the nearest-neighbor
carbon atoms, with the value $t_{0}=2.4$ eV, consistent with our
earlier calculations on conjugated polymers,\cite{PhysRevB.65.125204Shukla65,PhysRevB.69.165218Shukla69,PhysRevB.71.165204Priya_t0,:/content/aip/journal/jcp/131/1/10.1063/1.3159670Priyaanthracene,doi:10.1021/jp408535u,himanshu-triplet,sony-acene-lo}
polyaromatic hydrocarbons,\cite{doi:10.1021/jp410793rAryanpour,:/content/aip/journal/jcp/140/10/10.1063/1.4867363Aryanpour}
and graphene quantum dots.\cite{Tista1,Tista2} The next two terms
in Eq. \ref{eq:ppp} represent the electron-electron repulsion interactions:
(a) parameter $U$ denotes the on-site term, while (b) $V_{ij}$ denotes
the long-range Coulomb term. The distance-dependence of parameters
$V_{ij}$ is assumed as per Ohno relationship\cite{Theor.chim.act.2Ohno}

\begin{equation}
V_{ij}=U/\kappa_{i,j}(1+0.6117R_{i,j}^{2})^{\nicefrac{1}{2}},\label{eq:ohno}
\end{equation}

where $\kappa_{i,j}$ is the dielectric constant of the system, included
to take into account the screening effects, and $R_{i,j}$ is the
distance (in \AA) between the carbon atoms involved. In the present
set of calculations we have used two sets of Coulomb parameters: (a)
the ``screened parameters''\cite{PhysRevB.55.1497Chandross} with
$U=8.0$ eV, $\kappa_{i,j}=2.0(i\neq j)$, and $\kappa_{i,i}=1.0$,
and (b) the ``standard parameters'' with $U=11.13$ eV and $\kappa_{i,j}=1.0$. 

We initiate the computations by performing restricted Hartree-Fock
(RHF) calculations for the closed-shell singlet ground states of the
RGMs considered here, by employing the PPP Hamiltonian (Eq. \ref{eq:ppp}),
using a computer program developed in our group.\cite{Sony2010821}
The molecular orbitals (MOs) obtained from the RHF calculations are
used to transform the PPP Hamiltonian from the site basis, to the
MO basis, for the purpose of performing many-body calculations using
the CI approach. The correlated-electron multi-reference singles-doubles
configuration interaction (MRSDCI) approach was employed in this work,
which consists of a CI expansion obtained by exciting up to two electrons,
from a chosen list of reference configurations, to the unoccupied
MOs..\cite{peyerimhoff_energy_CI,buenker1978applicability} {The
reference configurations included in the MRSDCI method depend upon
the targeted states, which, in the present calculations are configurations
of symmetry $A_{g}$ for calculating the ground state, and configurations
of $B_{2u}$ and $B_{3u}$ symmetries for computing the one-photon
excited state wave functions.} {The MRSDCI calculations
are initiated using a single configuration, such as the closed-shell
RHF state, as the reference for the ground state, or a suitable set
of excited configurations of appropriate point-group symmetries, and
singlet spin multiplicity, for representing optically excited states
of the system. After performing the MRSDCI calculations with this
initial set of reference configurations, the optical absorption spectrum
is computed, and those excited excited states are identified which
contribute to peaks in it. Next, the wave functions of both ground
and excited states are carefully examined, and those configurations
are identified, magnitudes of whose coefficients are above a chosen
convergence threshold. The next MRSDCI calculation is carried out
with an enhanced reference set, obtained by including these additional
configurations. This procedure is iterated until the desired physical
quantities of the system, such as the excitation energies, optical
absorption spectra etc., converge. The lowest eigenvalue and eigenvector
obtained from the MRSDCI calculations on the $A_{g}$ symmetry manifold
corresponds to the ground state, while all other states are identified
with various excited states. For computing the $1^{3}B_{2u}$ state
needed for calculating the singlet-triplet splitting, the configurations
of triplet multiplicity, and $B_{2u}$ point-group symmetry, were
chosen. Thus, in all the MRSDCI calculations, only the configurations
consistent with the spin and point group symmetries of the targeted
states are included, making the calculations strictly spin- and symmetry
adapted, leading to tremendous computational savings. }

{In addition to the point-group and spin symmetries,
CI wave functions obtained in these calculations also possess electron-hole
(e-h) symmetry, which is a consequence of: (a) employing only nearest-neighbor
hoppings rendering the systems bipartite, and (b) all molecules considered
are half-filled, i.e., have one electron, per carbon atom. Conventionally,
the $1^{1}A_{g}$ ground state is assigned the negative (-) e-h parity,
while dipole selection rules require that the one-photon excited states
should possess not only opposite spatial parity ($u$, i.e., ungerade),
but also opposite e-h parity, i.e., positive (+) parity. Therefore,
all $B_{2u}/B_{3u}$ optically active excited states considered in
this work have positive (+) e-h parity.}

{In smaller RGMs, all the orbitals were treated as
active during the CI calculations. However, in cases of larger molecules
namely RGM-50, -54, and -56, we had to resort to the frozen orbital
approximation to keep the CI expansion tractable.} This consists of
freezing a few lowest energy occupied MOs, and removing the corresponding
symmetric virtual MOs from the list, as described in our earlier works.\cite{doi:10.1021/jp408535u,himanshu-triplet,Tista1}

Once CI calculations are finished, the many-body wave functions obtained
are used to compute the electric dipole transition matrix elements
connecting one-photon excited states to the ground state. The transition
dipole elements, along with the excitation energies of the excited
states, are used to compute the optical absorption cross-section $\sigma(\omega)$,
according to the formula 
\begin{equation}
\sigma(\omega)=4\pi\alpha\sum_{i}\frac{\omega_{i0}|\langle i|\boldsymbol{\hat{e.r}}|0\rangle|^{2}\gamma^{2}}{(\omega_{i0}-\omega)^{2}+\gamma^{2}}.\label{eq:sigma}
\end{equation}
In the equation above, $\omega$ denotes frequency of the incident
light, $\boldsymbol{\hat{e}}$ represents its polarization direction,
$\boldsymbol{r}$ is the position operator, $\alpha$ denotes the
fine structure constant, indices $0$ and $i$ represent, respectively,
the ground and excited states, $\omega_{i0}$ is the frequency difference
between those states, and $\gamma$ is the assumed universal line
width. The summation over $i$, in principle, is over an infinite
number of states which are dipole connected to the ground state. However,
in practice, the sum includes only those excited states whose excitation
energies are within a certain cutoff, normally taken to be 8 eV.

\section*{{\normalsize{} Results and Discussion}}

\label{sec:results}

In order to assess the role played by electron-correlation effects
on various properties of RGMs, it is important first to understand
the independent particle results obtained using the tight-binding
(TB) model. Therefore, in this section, we first present the result
obtained from the  TB model, followed by those obtained by the PPP-model. 

\subsection*{Tight-Binding Model Results}

\begin{figure}
\begin{centering}
\includegraphics[scale=0.6]{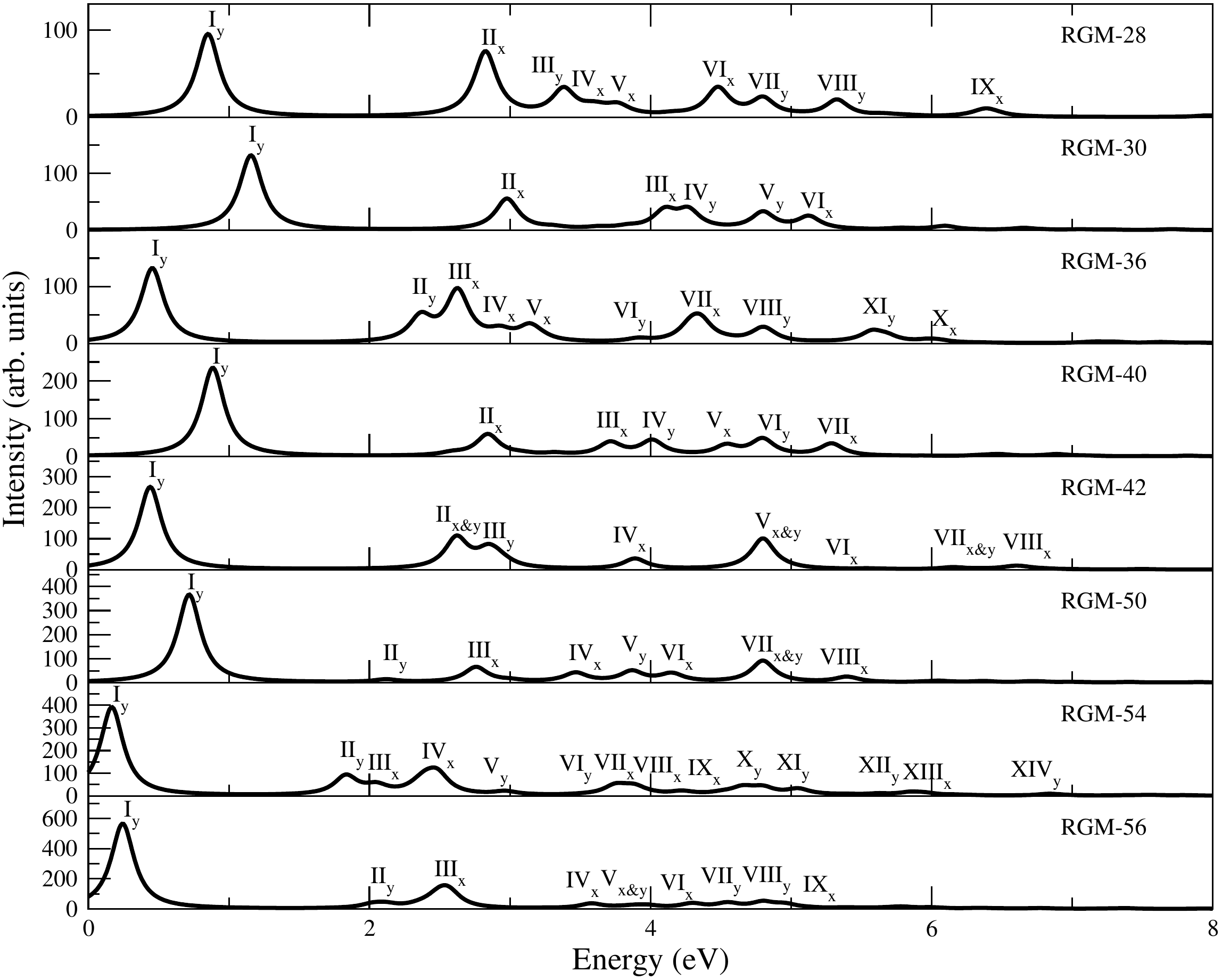}
\par\end{centering}
\centering{}\protect\caption{{\scriptsize{}\label{fig:Computed-linear-optical_TB}}Linear optical
absorption spectra for RGMs, computed using the TB model. The spectrum
has been broadened with a uniform line-width of 0.1 eV.}
\end{figure}

In Fig. \ref{fig:Computed-linear-optical_TB}, we have presented optical
absorption spectra of RGMs of varying sizes computed using the TB
model. The following conclusions can be drawn from this graph: 
\begin{enumerate}
\item The first peak in the absorption spectra for all the RGMs is $y-$polarized,
and corresponds to excitation of an electron from HOMO to LUMO, leading
to the excited state $1^{1}B_{2u}$. Therefore, this peak corresponds
to the optical gap of the concerned RGM, and it is the most intense
peak in the spectra. 
\item The intensity of the first peak in the spectra increases significantly
with the increasing size of RGMs. 
\item We also note that all $y-$polarized peaks are non-degenerate, and
correspond to excited states of symmetry $^{1}B_{2u}$, as per the
selection rules of $D_{2h}$ point group. All $x-$polarized peaks
are doubly degenerate, as a consequence of electron-hole symmetry
of the nearest neighbor TB model, and correspond to excited states
of $^{1}B_{3u}$ symmetry. For example, in RGM-30, first peak is $y-$polarized,
and it is due to non-degenerate excitation $\left|H\rightarrow L\right\rangle $,
while the second peak is $x-$polarized, and is due to doubly degenerate
excitations $\left|H\rightarrow L+3\right\rangle $ and $\left|H-3\rightarrow L\right\rangle $.
Notations $H/L$ imply HOMO/LUMO, while $H-m$ ($L+n)$ imply $m$-th
orbital below HOMO ($n$-th orbital above LUMO).
\item With the increasing length of the RGMs along a given orientation (zigzag
or armchair), the optical gap decreases. 
\end{enumerate}
In Table \ref{tab:optical gap-(H-L)-gap_RGQD}, we compare the HOMO-LUMO
gap obtained at the tight-binding level, with the experimental results,
wherever available. For the sake of comparison, we also present the
values of optical gaps obtained using the PPP-CI approach to be discussed
in the next section. From the table it is obvious that the gaps obtained
using the TB model are much smaller than the experimental values.
On the other hand, the PPP-CI values of the optical gaps, are generally
in much better agreement with the experiments. Therefore, it is obvious
that the TB model cannot provide good quantitative agreement with
the experiments, because it ignores the electron-correlation effects. 

\begin{table}
\caption{\label{tab:optical gap-(H-L)-gap_RGQD}Optical gaps of various RGMs
obtained using the TB model, and the PPP model. In case of PPP model,
the gaps are computed using the CI approach, by employing both the
screened and the standard parameters, denoted  as Scr, and Std,
respectively. }

\centering{}%
\begin{tabular}{|c|c|c|c|c|}
\hline 
\multirow{3}{*}{System} & \multicolumn{4}{c|}{optical gap (eV)}\tabularnewline
\cline{2-5} 
 & \multirow{2}{*}{TB Model} & \multicolumn{2}{c|}{PPP-CI} & \multirow{2}{*}{Experimental}\tabularnewline
\cline{3-4} 
 &  & Scr & Std & \tabularnewline
\hline 
RGM-28 & 0.85 & 2.00 & 2.21 & 1.80\cite{Konishi_Bisanthenes}, 2.02\cite{clar1977corr}, 2.15\cite{arabei2000} \tabularnewline
\hline 
\multirow{2}{*}{RGM-30} & \multirow{2}{*}{1.16} & \multirow{2}{*}{2.11} & \multirow{2}{*}{2.43} & 2.14\cite{kummer1997absorption}, 2.21\cite{baumgarten1994spin,biktchantaev2002perylene},
2.22\cite{koch1991polyarylenes}, \tabularnewline
 &  &  &  & 2.35\cite{ruiterkamp2002spectroscopy}, 2.36\cite{clar1978correlations,halasinski2003electronic} \tabularnewline
\hline 
RGM-36 & 0.45 & 2.11 & 2.30 & -\tabularnewline
\hline 
\multirow{2}{*}{RGM-40} & \multirow{2}{*}{0.89} & \multirow{2}{*}{2.02} & \multirow{2}{*}{2.30} & 1.84\cite{former2002cyclodehydrogenation}, 1.87\cite{koch1991polyarylenes,baumgarten1994spin},
1.91\cite{gudipati2006double}, \tabularnewline
\cline{5-5} 
 &  &  &  & 2.03\cite{clar1978correlations}, 2.04\cite{ruiterkamp2002spectroscopy,halasinski2003electronic} \tabularnewline
\hline 
RGM-42 & 0.44 & 1.86 & 2.04 & -\tabularnewline
\hline 
RGM-50 & 0.72 & 1.72 & 1.98 & 1.66\cite{baumgarten1994spin,koch1991polyarylenes} \tabularnewline
\hline 
RGM-54 & 0.17 & 1.63 & 2.09 & -\tabularnewline
\hline 
RGM-56 & 0.24 & 1.50 & 1.91 & 1.35\cite{konishi2014anthenes} \tabularnewline
\hline 
\end{tabular}
\end{table}

\subsection*{PPP Model Based CI Results}

In this section, we present our results obtained from the PPP-model
based CI calculations. First, we present the results on the nature
of ground state wave function of RGMs, followed by their spin gaps.
Finally, we present and discuss the calculated linear optical absorption
spectra of these molecules.

\subsubsection*{Nature of Ground State}

Before discussing the nature of ground state wave functions of RGMs,
we would like to make a brief general comment about the number of
active orbitals involved in the MRSDCI calculations performed on these
molecules. For RGM-$n$, $n=28$ to $42$, all the $n$ MOs of the
systems were included in the MRSDCI calculations. However, from $n=50$
to $72$, it was no longer possible to perform accurate calculations
with all the $n$ MOs treated as active. Therefore, for these cases,
some low-lying occupied MO's, were frozen, while their unoccupied
counterparts were deleted. As a result, the number of active orbitals
($N_{act}$) for $n=50-56$ was 42, while for $n=72$, it was 44.
The final choice of $N_{act}$ for each molecule was decided after
careful convergence considerations both for the singlet-triplet gap,
and the optical absorption spectra. For $n=50$, we have explicitly
presented the convergence of optical absorption spectra with respect
to $N_{act}$.

Because, to compute the spin gaps we needed energies of $1^{1}A_{g}$
and $1^{3}B_{2u}$ states which are in different symmetry manifolds,
we managed to perform {reasonably large} MRSDCI calculations,
as is evident from Table S1 of Supporting Information. Therefore,
we believe that the wave functions and the energies of the ground
state ( $1^{1}A_{g}$ ), and the lowest triplet state $1^{3}B_{2u}$
are very accurate.

\begin{table}
\caption{\label{tab:WF-of-ground-state}Configurations making significant contributions
to the ground state ($1^{1}A_{g}$ ) wave functions of RGM-$n$ ($n=28-72$),
computed using the MRSDCI approach, and the standard (Std) and screened
(Scr) parameters in the PPP-model Hamiltonian. $|HF\rangle$
denotes the closed-shell restricted Hartree-Fock configuration, with
respect to which other configurations are defined. In particular,
$|H\rightarrow L;H\rightarrow L\rangle$ denotes the doubly-excited
configuration with respect to the $|HF\rangle$, obtained by promoting
two electrons from HOMO ($H$) to LUMO ($L$), of the concerned RGM.
The expansion coefficient of each configuration in the ground state
wave function is written in the parenthesis next to it. }

\centering{}%
\begin{tabular}{|c|c|c|}
\hline 
\multirow{1}{*}{system} & Scr & Std\tabularnewline
\hline 
\multirow{2}{*}{RGM-28} & $|HF\rangle$(0.7899) & $|HF\rangle$(0.8315 )\tabularnewline
\cline{2-3} 
 & $|H\rightarrow L;H\rightarrow L\rangle$ (0.2264) & $|H\rightarrow L;H\rightarrow L\rangle$ (0.2362)\tabularnewline
\hline 
\multirow{2}{*}{RGM-30} & $|HF\rangle$(0.8143) & $|HF\rangle$(0.8574)\tabularnewline
\cline{2-3} 
 & $|H\rightarrow L;H\rightarrow L\rangle$ (0.1685) & $|H\rightarrow L;H\rightarrow L\rangle$ (0.1415)\tabularnewline
\hline 
\multirow{2}{*}{RGM-36} & $|HF\rangle$(0.7269) & $|HF\rangle$(0.7572)\tabularnewline
\cline{2-3} 
 & $|H\rightarrow L;H\rightarrow L\rangle$ (0.3468) & $|H\rightarrow L;H\rightarrow L\rangle$ (0.3399)\tabularnewline
\hline 
\multirow{2}{*}{RGM-40} & $|HF\rangle$(0.7946) & $|HF\rangle$(0.8401)\tabularnewline
\cline{2-3} 
 & $|H\rightarrow L;H\rightarrow L\rangle$ (0.1796) & $|H\rightarrow L;H\rightarrow L\rangle$ (0.1534)\tabularnewline
\hline 
\multirow{2}{*}{RGM-42} & $|HF\rangle$(0.7247) & $|HF\rangle$(0.7530)\tabularnewline
\cline{2-3} 
 & $|H\rightarrow L;H\rightarrow L\rangle$ (0.3508) & $|H\rightarrow L;H\rightarrow L\rangle$ (0.3342)\tabularnewline
\hline 
\multirow{2}{*}{RGM-50} & $|HF\rangle$(0.8044) & $|HF\rangle$(0.8434)\tabularnewline
\cline{2-3} 
 & $|H\rightarrow L;H\rightarrow L\rangle$ (0.2064) & $|H\rightarrow L;H\rightarrow L\rangle$ (0.1586)\tabularnewline
\hline 
\multirow{2}{*}{RGM-54} & $|HF\rangle$(0.6504) & $|HF\rangle$(0.6400)\tabularnewline
\cline{2-3} 
 & $|H\rightarrow L;H\rightarrow L\rangle$ (0.4866) & $|H\rightarrow L;H\rightarrow L\rangle$ (0.4975)\tabularnewline
\hline 
\multirow{2}{*}{RGM-56} & $|HF\rangle$(0.6798) & $|HF\rangle$(0.6724 )\tabularnewline
\cline{2-3} 
 & $|H\rightarrow L;H\rightarrow L\rangle$ (0.4541) & $|H\rightarrow L;H\rightarrow L\rangle$ (0.4469)\tabularnewline
\hline 
\multirow{2}{*}{RGM-72} & $|HF\rangle$(0.6164) & $|HF\rangle$(0.5958)\tabularnewline
\cline{2-3} 
 & $|H\rightarrow L;H\rightarrow L\rangle$ (0.5562) & $|H\rightarrow L;H\rightarrow L\rangle$ (0.5666)\tabularnewline
\hline 
\end{tabular}
\end{table}
The dominant electronic configurations contributing to the ground
state MRSDCI wave functions various RGMs are presented in Table \ref{tab:WF-of-ground-state}.
An inspection of the table reveals the following trends: (a) In all
cases, the most dominant configuration to the ground state wave function
is the closed-shell RHF configuration, (b) The relative magnitude
of the RHF configuration to the wave function decreases significantly
with the increasing sizes of RGMs. This decreases is accompanied by
a significant increase in the relative contribution of the doubly-excited
configuration $|H\rightarrow L;H\rightarrow L\rangle$ to the wave
function. For example, for the smallest molecule RGM-28 the coefficients
of $|HF\rangle$ and $|H\rightarrow L;H\rightarrow L\rangle$ are
close to 0.80 and 0.20, respectively, while for the largest one RGM-72,
they are 0.60 and 0.56, respectively, i.e., almost equal. Because
these trends are true, irrespective of the Coulomb parameters used
in the calculations, it is obvious that the ground states of RGMs
are developing a significantly open-shell diradical character with
the increasing sizes. This result was also observed in our earlier
work for oligoacenes of increasing sizes, although the extent of configuration
mixing for acenes was much smaller compared to RGMs.\cite{doi:10.1021/jp408535u}
{Plasser et al.\cite{plasser2013multiradical} performed
first-principles multi-reference averaged quadratic coupled cluster
(MR-AQCC) calculations on the ground states of a large number of finite
hydrogen-saturated graphene-like molecules, and computed the natural
orbital (NO) occupancies of the wave functions. The molecules common
between our calculations and those of Plasser et al.\cite{plasser2013multiradical}
are RGM-30, RGM-42, and RGM-54. They also concluded that RGMs have
a significant open-shell character. In Table S3 of the Supporting
Information, we present the numbers of singly- and doubly-occupied
orbitals of various irreducible representations, based upon the configurations
whose coefficient are larger than 0.05 in the ground-state MRSDCI
wave functions of various RGMs. These numbers are in good agreement
with the number of NOs with occupancies close to one and two, reported
by Plasser }{\emph{et al}}{.\cite{plasser2013multiradical}
for RGM-30, RGM-42, and RGM-54.}

\subsubsection*{Spin Gaps}

Spin gap of an electronic system is the energy difference between
the lowest triplet and singlet states. For RGMs, the lowest singlet
state is $1^{1}A_{g}$ ground state, while the lowest triplet state
is the $1^{3}B_{2u}$ state, whose spatial part of the wave function
consists predominantly of the single excitation $|H\rightarrow L\rangle$,
just as in the case of $1^{1}B_{2u}$ state. Thus, at the TB level,
$1^{1}B_{2u}$and $1^{3}B_{2u}$ will be degenerate, and, therefore
their spin and optical gaps will be identical. However, if the two
gaps are found to be different for RGMs, it can only be due to electron-electron
interactions. Therefore, difference in the spin and optical gaps is
a measure of the electron correlation effects in RGMs. With this in
mind, we computed the spin gaps of RGMs up to RGM-72, using our PPP
model based MRSDCI approach. Given that the size
of these CI calculations (see Table S1 of Supporting Information)
is {reasonably large}, we believe that the spin gaps
of RGMs presented in Table \ref{tab:Spin-gap} are fairly accurate.
{Furthermore, in the same table, our results are compared
with those reported by Horn $et.$ $al$. \cite{horn2016comparison},
obtained from first-principles electron-correlated calculations performed
on RGM-30, RGM-42, and RGM-54. It is obvious that the results of Horn
}{\emph{et al}.\cite{horn2016comparison},
obtained using the $\pi$-electron MR-AQCC approach, are in good agreement
with our results.

From the inspection of Table \ref{tab:Spin-gap} it is obvious that
spin gaps of RGMs are decreasing with their increasing sizes. Thus,
this result suggests that the spin gap of infinite graphene is zero,
consistent with the widespread assumption that graphene is a weakly-correlated
material. It is noteworthy that based on an identical PPP-MRSDCI methodology,
in an earlier work from our group, it was shown that the spin gaps
of oligoacenes exhibit signs of saturation with the increasing conjugation
length,\cite{doi:10.1021/jp408535u} indicating stronger electron-correlation
effects. Given the fact that the oligoacenes are nothing but hydrogen-passivated,
finite-sized, narrowest possible ($N_{z}=2)$ zigzag nanoribbons,
this suggests that electron-correlation effects are stronger in them,
as compared to graphene, because of their reduced dimensionality.

We present the important configurations contributing to the many body
wave functions of $1^{3}B_{2u}$ states of various RGMs in Table \ref{tab:WF-of-lowest-energy-states-1}.
It is obvious from the table that although the single excitation $|H\rightarrow L\rangle$
makes the dominant contribution to the triplet wave function in all
the cases, but other configurations also make smaller, but, significant
contributions, thereby underlying the importance of including electron-correlation
effects in accurate quantitative calculation of energies of triplet
states.

It will also be interesting to compare the experimental values of
the spin-gaps of individual hydrocarbon molecules corresponding to
these RGMs, with our calculated values. Therefore, we urge the experimentalists
to measure the spin gaps of RGMs studied in the present work.

\begin{table}
\caption{\label{tab:Spin-gap} Singlet-Triplet gaps ($\Delta E_{ST}=E(1^{3}B_{2u})-E(1^{1}A_{g}))$
of RGMs, computed using the MRSDCI method, employing screened (Scr)
and standard parameters (Std) in the PPP model. RGMs are divided in
groups of three, where each group corresponds to a common width, and
increasing armchair length.}

\begin{centering}
\begin{tabular}{|c|c|c|c|}
\hline 
\multirow{2}{*}{System} & \multicolumn{2}{c|}{$\Delta E_{ST}$ (eV)} & {$\Delta E_{ST}$ (eV)}\tabularnewline
\cline{2-4} 
 & Scr & Std & {Theory (others)\cite{horn2016comparison} }\tabularnewline
\hline 
RGM-30 & 1.11 & 1.30 & {1.10$^{a}$, 1.68$^{b}$, 1.95$^{c}$, 2.31$^{d}$}\tabularnewline
RGM-40 & 0.97 & 1.16 & {-}\tabularnewline
RGM-50 & 0.79 & 0.94 & {-}\tabularnewline
\hline 
RGM-28 & 0.76 & 0.75 & {-}\tabularnewline
RGM-42 & 0.40 & 0.36 & {0.30$^{a}$, 0.26$^{b}$, 0.23$^{c}$, 0.21$^{d}$}\tabularnewline
RGM-56 & 0.15 & 0.11 & {-}\tabularnewline
\hline 
RGM-36 & 0.37 & 0.34 & {-}\tabularnewline
RGM-54 & 0.13 & 0.07 & {0.05$^{a}$, 0.04$^{b}$, 0.05$^{c}$, 0.07$^{d}$}\tabularnewline
RGM-72 & 0.06 & 0.03 & {-}\tabularnewline
\hline 
\multicolumn{4}{|c|}{{$\pi$-MR-AQCC$^{a}$, $\pi$-MR-CISD+Q$^{b}$, $\pi$-MR-CISD$^{c}$,
$\pi$-MCSCF$^{d}$}}\tabularnewline
\hline 
\end{tabular}
\par\end{centering}
\protect
\end{table}
\begin{table}
\caption{\label{tab:WF-of-lowest-energy-states-1}Configurations making significant
contributions to the lowest triplet state ($1^{3}B_{2u}$ ) wave functions
of RGM-$n$ ($n=28-72$), computed using the MRSDCI approach, and
the standard (Std) and screened (Scr) parameters in the PPP-model
Hamiltonian. Various configurations  are defined with
respect to the closed-shell restricted Hartree-Fock configuration
$|HF\rangle$. $|H\rightarrow L\rangle$ denotes the singly-excited
configuration with respect to the $|HF\rangle$, obtained by promoting
one electron from HOMO ($H$) to LUMO ($L$), of the concerned RGM.
The expansion coefficient of each configuration in the ground state
wave function is written in the parenthesis next to it. }

\centering{}%
\begin{tabular}{|c|c|c|}
\hline 
\multirow{1}{*}{system} & Scr & Std\tabularnewline
\hline 
\multirow{2}{*}{RGM-28} & $|H\rightarrow L\rangle$ (0.8097 ) & $|H\rightarrow L\rangle$ (0.8197)\tabularnewline
\cline{2-3} 
 & $|H-1\rightarrow L+1\rangle$ (0.1390) & $|H\rightarrow L;H-2\rightarrow L\rangle+c.c.$ (0.1480)\tabularnewline
\hline 
\multirow{2}{*}{RGM-30} & $|H\rightarrow L\rangle$ (0.8035) & $|H\rightarrow L\rangle$ (0.8049)\tabularnewline
\cline{2-3} 
 & $|H-1\rightarrow L+1\rangle$ (0.1545) & $|H-1\rightarrow L+1\rangle$ (0.1685)\tabularnewline
\hline 
\multirow{2}{*}{RGM-36} & $|H\rightarrow L\rangle$ (0.8047) & $|H\rightarrow L\rangle$ (0.8022)\tabularnewline
\cline{2-3} 
 & $|H-1\rightarrow L+1\rangle$ (0.1247) & $|H\rightarrow L;H-2\rightarrow L\rangle-c.c.$ (0.1522)\tabularnewline
\hline 
\multirow{2}{*}{RGM-40} & $|H\rightarrow L\rangle$ (0.7912) & $|H\rightarrow L\rangle$ (0.7876)\tabularnewline
\cline{2-3} 
 & $|H-1\rightarrow L+1\rangle$ (0.1681) & $|H-1\rightarrow L+1\rangle$ (0.1869)\tabularnewline
\hline 
\multirow{2}{*}{RGM-42} & $|H\rightarrow L\rangle$ (0.8055) & $|H\rightarrow L\rangle$ (0.7838)\tabularnewline
\cline{2-3} 
 & $|H\rightarrow L;H-2\rightarrow L\rangle+c.c.$ (0.1312) & $|H\rightarrow L;H-2\rightarrow L\rangle-c.c.$ (0.1922)\tabularnewline
\hline 
\multirow{2}{*}{RGM-50} & $|H\rightarrow L\rangle$ (0.8060) & $|H\rightarrow L\rangle$ (0.7785)\tabularnewline
\cline{2-3} 
 & $|H-1\rightarrow L+1\rangle$ (0.1224) & $|H-1\rightarrow L+1\rangle$ (0.1974)\tabularnewline
\hline 
\multirow{2}{*}{RGM-54} & $|H\rightarrow L\rangle$ (0.8188) & $|H\rightarrow L\rangle$ (0.7984)\tabularnewline
\cline{2-3} 
 & $|H\rightarrow L;H-2\rightarrow L\rangle-c.c.$(0.1173) & $|H\rightarrow L;H-2\rightarrow L\rangle-c.c.$(0.1889)\tabularnewline
\hline 
\multirow{2}{*}{RGM-56} & $|H\rightarrow L\rangle$ (0.8127) & $|H\rightarrow L\rangle$ (0.7697)\tabularnewline
\cline{2-3} 
 & $|H\rightarrow L;H-1\rightarrow L\rangle-c.c.$ (0.1496) & $|H\rightarrow L;H-1\rightarrow L\rangle-c.c.$(0.2295)\tabularnewline
\hline 
\multirow{2}{*}{RGM-72} & $|H\rightarrow L\rangle$ (0.8417) & $|H\rightarrow L\rangle$ (0.8196)\tabularnewline
\cline{2-3} 
 & $|H\rightarrow L;H-2\rightarrow L\rangle+c.c.$(0.1111) & $|H\rightarrow L;H-2\rightarrow L\rangle-c.c.$(0.1738)\tabularnewline
\hline 
\end{tabular}
\end{table}

\subsubsection*{Linear optical absorption spectrum}

In this section, we present optical absorption spectra of RGM-$n$,
with $n$ ranging from 30 to 56, computed using the PPP model, and
the MRSDCI approach. Before discussing the results of our calculations,
in Table S2 of Supporting Information we give the sizes of the CI
matrices, for different symmetry spaces of various RGMs. The fact
that the sizes of the CI matrices were in the range $2.43\times10^{5}-6.19\times10^{6}$,
implies that these calculations were {reasonably large},
and, therefore, should be fairly accurate. 

The calculated spectra of these RGMs are presented in Figs. \ref{fig:Computed-linear-optical_ppp-CI_scr} and \ref{fig:Computed-linear-optical_ppp-CI_std}, while the important information regarding the excited states contributing
to various peaks in the spectra, including their wave functions, are
presented in Tables S12-S27 of Supporting Information.

\begin{figure}
\begin{centering}
\includegraphics[scale=0.5]{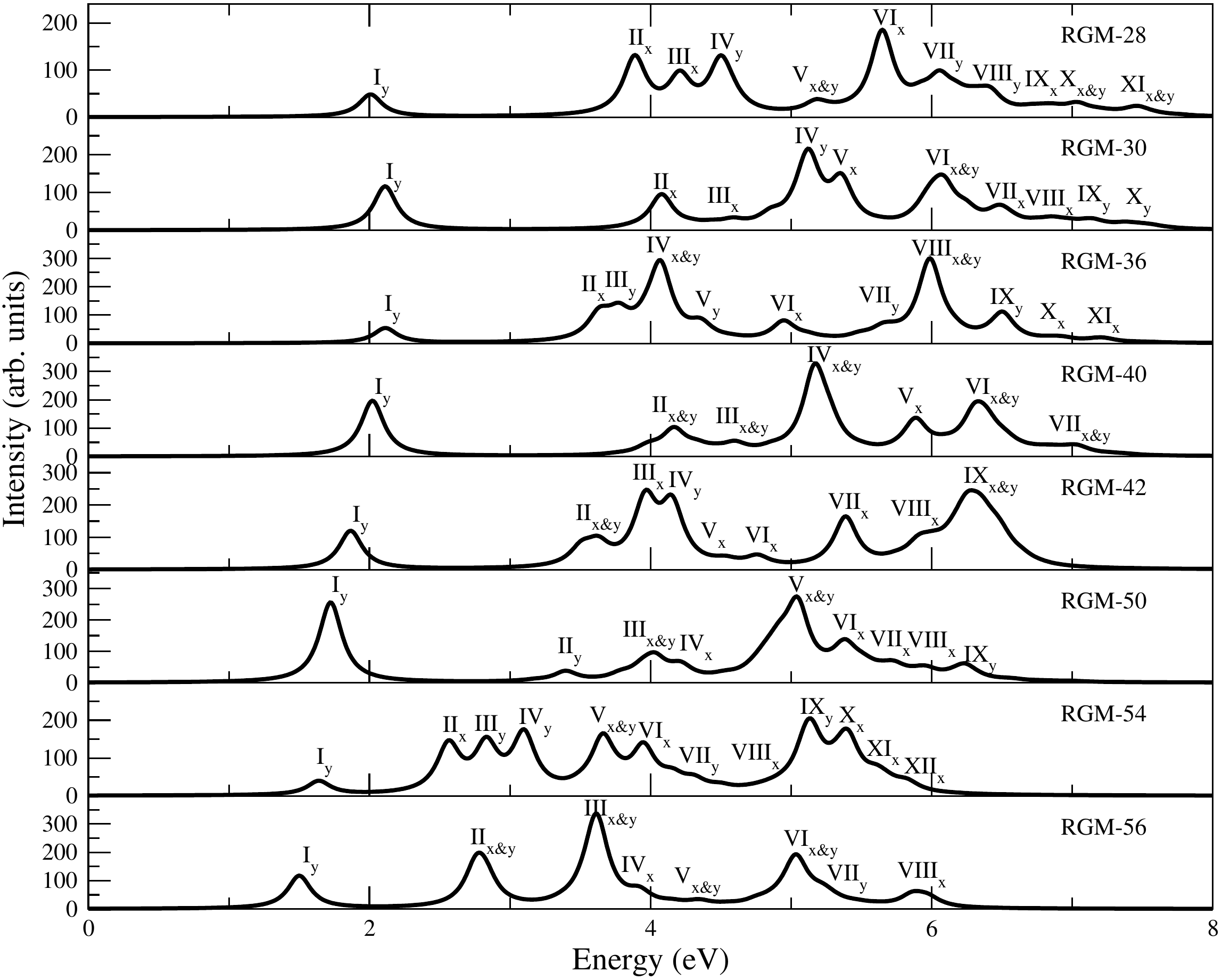}
\par\end{centering}

\centering{}\caption{{\scriptsize{}\label{fig:Computed-linear-optical_ppp-CI_scr}}Computed
linear optical absorption spectra of RGMs, obtained using the MRSDCI
approach, by employing screened Coulomb parameters in the PPP
model. The spectra have been broadened using a uniform line-width of 0.1 eV.}
\end{figure}

\begin{figure}
\begin{centering}
\includegraphics[scale=0.5]{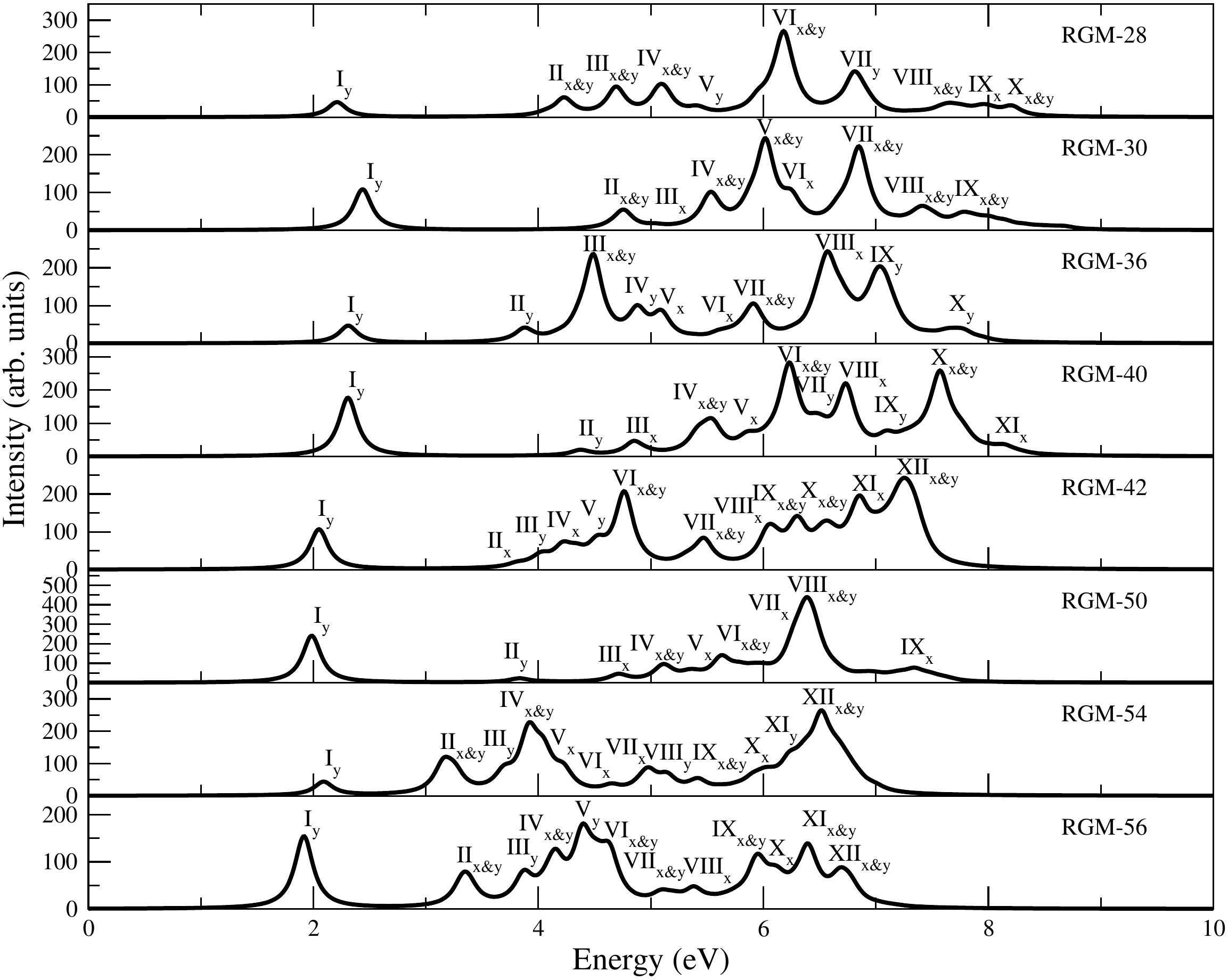}
\par\end{centering}

\centering{}\caption{{\scriptsize{}\label{fig:Computed-linear-optical_ppp-CI_std}}Computed
linear optical absorption spectra of RGMs, obtained using the MRSDCI
approach, by employing standard Coulomb parameters in the PPP model.
The spectra have been broadened using a uniform
line-width of 0.1 eV.}
\end{figure}

Before discussing the spectra of individual RGMs, we discuss the general
trends observed in our calculation: 
\begin{enumerate}
\item For each RGM-$n$, the absorption spectrum obtained using the PPP-CI
approach is blueshifted in comparison to the TB model.
\item For all the RGMs, absorption spectra obtained using the screened parameters
are red-shifted compared to those obtained using the standard parameters.
\item In all cases, the first peak of the spectrum is due to optical excitation
from the $1^{1}A_{g}$ ground state to $1^{1}B_{2u}$ excited state,
and corresponds to the optical gap. As per electric dipole selection
rules, this peak is $y$-polarized. {The wave function
of the $1^{1}B_{2u}$ state for all the RGMs is dominated by singly-excited
configuration $|H\rightarrow L\rangle$, where $H$ and $L$, respectively,
denote the HOMO and the LUMO of the system. }
\item For all the RGMs, the first peak is not the most intense peak. For
a number of RGMs, several high energy peaks are more intense than
the first one. This result is in sharp contrast with the TB model
results.
\item Dominant configurations in the wave functions of the excited states
corresponding to the lower energy peaks are single excitations, while
those in the higher energy peaks are dominated by double and higher
excitations.
\end{enumerate}
{In Table \ref{tab:Comparison-RGM}, we present the
locations of the peaks corresponding to the optical gap, and a higher
energy $^{1}B_{3u}$ state with dominant contribution to the oscillator
strength, obtained from our calculations. In the same table, for the
sake of comparison, we also present the corresponding experimental
results, and the theoretical calculations of other authors, for all
the RGMs considered in this work. It is obvious from the table that
the agreement between the results of our calculations, and the experimental
ones, is quite good for all the RGMs. Additionally, a detailed comparison
for all the important peaks of individual RGMs is presented in Tables
 S4-S11 of the Supporting Information.} For RGM-54, we could not locate
any previous experimental or theoretical data. Next, we discuss our
results for the individual RGMs.

\begin{table}
\caption{\label{tab:Comparison-RGM}Comparison of our calculations with the
experiments, and the theoretical works of other authors, for the peaks
corresponding to: (a) optical gap ($1^{1}B_{2u}$state), and (b) a
higher energy peak with the dominant contribution from a $^{1}B_{3u}$
state, for various RGMs. Our calculations were performed using the
PPP-MRSDCI approach, employing both the screened (Scr.) and the standard
(Std.) parameters. All results are in eV units.}
\begin{tabular}{|c|c|c|c|c|}
\hline 
\multicolumn{5}{|c|}{Excitation energy ($eV$)}\tabularnewline
\hline 
\hline 
\multirow{2}{*}{RGM} & \multicolumn{2}{c|}{This work} & \multirow{2}{*}{Experimental} & \multirow{2}{*}{Other theoretical}\tabularnewline
\cline{2-3} 
 & Scr. (Peak) & Std. (Peak) &  & \tabularnewline
\hline 
\multirow{4}{*}{28} & 2.00 ($I_{y}$) & 2.21 ($I_{y}$) & 1.80\cite{Konishi_Bisanthenes},1.98\cite{Konishi_Bisanthenes},
2.02\cite{clar1977corr}, & \href{http://www.dsf.unica.it/~gmalloci/pahs/bisanthene/bisanthene.html}{1.47}\cite{online_RQD28}\tabularnewline
 & $(^{1}B_{2u})$ & $(^{1}B_{2u})$ & 2.15\cite{arabei2000},2.43\cite{arabei2000}, & 1.78$^{a}$\cite{parac2003tddft}, 1.98$^{b}$\cite{parac2003tddft}\tabularnewline
\cline{2-5} 
 & 4.19 ($III_{x}$) & 4.14 ($II_{x\&y}$) & \multirow{2}{*}{4.05\cite{Konishi_Bisanthenes}} & \multirow{2}{*}{-}\tabularnewline
 & $(^{1}B_{3u})$ & $(^{1}B_{2u}/^{1}B_{3u})$ &  & \tabularnewline
\hline 
\multirow{4}{*}{30} & 2.11 ($I_{y}$) & 2.43 ($I_{y}$) & 2.14\cite{kummer1997absorption},2.21\cite{baumgarten1994spin,biktchantaev2002perylene},
2.22\cite{koch1991polyarylenes},2.35\cite{ruiterkamp2002spectroscopy}, & 2.02\cite{halasinski2003electronic},2.03\cite{viruela1992theoretical},2.21$^{a}$/2.22$^{c}$\cite{malloci2011electronic},
2.29\cite{minami2012theoretical},\tabularnewline
 & $(^{1}B_{2u})$ & $(^{1}B_{2u})$ & 2.36\cite{clar1978correlations,halasinski2003electronic},2.39\cite{koch1991polyarylenes},2.57\cite{koch1991polyarylenes},
2.76\cite{koch1991polyarylenes}, & 2.52\cite{karabunarliev1994structure}, 2.98\cite{halasinski2003electronic},
3.31\cite{halasinski2003electronic}, 3.40\cite{halasinski2003electronic},3.84\cite{halasinski2003electronic},\tabularnewline
\cline{2-5} 
 & 5.35 ($V_{x}$) & 5.53 ($IV_{x\&y}$) & 5.20\cite{halasinski2003electronic}, 5.27\cite{ruiterkamp2002spectroscopy},
5.41\cite{halasinski2003electronic}, 5.48\cite{koch1991polyarylenes} & -\tabularnewline
 & $(^{1}B_{3u})$ & $(^{1}B_{2u}/^{1}B_{3u})$ &  & \tabularnewline
\hline 
\multirow{4}{*}{36} & 2.11 ($I_{y}$) & 2.30 ($I_{y}$) & \multirow{2}{*}{-} & \multirow{2}{*}{-}\tabularnewline
 & $(^{1}B_{2u})$ & $(^{1}B_{2u})$ &  & \tabularnewline
\cline{2-5} 
 & 3.63 ($II_{x}$) & 3.87 ($II_{y}$) & \multirow{2}{*}{-} & \multirow{2}{*}{3.64\cite{online_RQD36}}\tabularnewline
 & $(^{1}B_{3u})$ & $(^{1}B_{2u})$ &  & \tabularnewline
\hline 
\multirow{4}{*}{40} & 2.02 ($I_{y}$) & 2.30 ($I_{y}$) & 1.84\cite{former2002cyclodehydrogenation}, 1.87\cite{baumgarten1994spin,koch1991polyarylenes},
1.91\cite{gudipati2006double}, & 1.65\cite{viruela1992theoretical},1.67\cite{halasinski2003electronic},1.79$^{c}$/1.83$^{a}$\cite{malloci2011electronic},
1.87\cite{gudipati2006double},\tabularnewline
 & $(^{1}B_{2u})$ & $(^{1}B_{2u})$ & 1.99\cite{former2002cyclodehydrogenation}, 2.03\cite{clar1978correlations},
2.04\cite{ruiterkamp2002spectroscopy,halasinski2003electronic}, & 1.88\cite{minami2012theoretical}, 2.18\cite{karabunarliev1994structure},
2.97\cite{halasinski2003electronic},\tabularnewline
\cline{2-5} 
 & 5.16 ($IV_{x\&y}$) & 5.48 ($IV_{x\&y}$) & \multirow{2}{*}{5.27\cite{koch1991polyarylenes}, 5.39 \cite{halasinski2003electronic}} & \multirow{2}{*}{5.30\cite{malloci2011electronic}}\tabularnewline
 & $(^{1}B_{2u}/^{1}B_{3u})$ & $(^{1}B_{2u}/^{1}B_{3u})$ &  & \tabularnewline
\hline 
\multirow{4}{*}{42} & 1.86 ($I_{y}$) & 2.04 ($I_{y}$) & \multirow{2}{*}{1.41\cite{konishi2014anthenes}, 1.57\cite{konishi2014anthenes}} & \multirow{2}{*}{-}\tabularnewline
 & $(^{1}B_{2u})$ & $(^{1}B_{2u})$ &  & \tabularnewline
\cline{2-5} 
 & 3.96 ($III_{x}$) & 3.80 $(II_{x})$ & \multirow{2}{*}{3.87\cite{konishi2014anthenes}} & \multirow{2}{*}{-}\tabularnewline
 & $(^{1}B_{3u})$ & $(^{1}B_{3u})$ &  & \tabularnewline
\hline 
\multirow{4}{*}{50} & 1.72 ($I_{y}$) & 1.98 ($I_{y}$) & \multirow{2}{*}{1.66\cite{baumgarten1994spin,koch1991polyarylenes}} & \multirow{2}{*}{1.40\cite{viruela1992theoretical},1.51$^{c}$/1.54$^{a}$\cite{malloci2011electronic},
1.60\cite{minami2012theoretical},1.97\cite{karabunarliev1994structure}}\tabularnewline
 & $(^{1}B_{2u})$ & $(^{1}B_{2u})$ &  & \tabularnewline
\cline{2-5} 
 & 4.97 ($V_{x\&y}$) & 5.12 ($IV_{x\&y}$) & \multirow{2}{*}{4.80\cite{koch1991polyarylenes}} & \multirow{2}{*}{5.2\cite{malloci2011electronic}}\tabularnewline
 & $(^{1}B_{2u}/{}^{1}B_{3u})$ & $(^{1}B_{2u}/{}^{1}B_{3u})$ &  & \tabularnewline
\hline 
\multirow{4}{*}{54} & 1.63 ($I_{y}$) & 2.09 ($I_{y}$) & \multirow{2}{*}{-} & \multirow{2}{*}{-}\tabularnewline
 & $(^{1}B_{2u})$ & $(^{1}B_{2u})$ &  & \tabularnewline
\cline{2-5} 
 & 2.56 ($II_{x}$) & 3.20 ($II_{x\&y}$) & \multirow{2}{*}{-} & \multirow{2}{*}{-}\tabularnewline
 & $(^{1}B_{3u})$ & $(^{1}B_{2u}/{}^{1}B_{3u})$ &  & \tabularnewline
\hline 
\multirow{4}{*}{56} & 1.50 ($I_{y}$) & 1.91 ($I_{y}$) & \multirow{2}{*}{1.35\cite{konishi2014anthenes}, 2.01\cite{konishi2014anthenes},
2.10\cite{konishi2014anthenes},} & \multirow{2}{*}{-}\tabularnewline
 & $(^{1}B_{2u})$ & $(^{1}B_{2u})$ &  & \tabularnewline
\cline{2-5} 
 & 2.79 ($II_{x\&y}$) & 3.35 ($II_{x\&y}$) & \multirow{2}{*}{3.21\cite{konishi2014anthenes}} & \multirow{2}{*}{-}\tabularnewline
 & $(^{1}B_{2u}/{}^{1}B_{3u})$ & $(^{1}B_{2u}/{}^{1}B_{3u})$ &  & \tabularnewline
\hline 
\multicolumn{5}{|c|}{$^{a}$ TDDFT method, $^{b}$ TDPPP method, $^{c}$DFT(Kohan-Sham)
method}\tabularnewline
\hline 
\end{tabular}
\end{table}

\subsubsection*{RGM-28}

Clar and Schmidt\cite{clar1977corr}, Arabei $et$ $al.$\cite{arabei2000},
and Konishi $et$ $al.$\cite{Konishi_Bisanthenes} have reported
the measurements of the absorption spectrum of bisanthene, and its
derivatives, the structural analogs of RGM-$28$. In Figs. \ref{fig:Computed-linear-optical_ppp-CI_scr}
 and \ref{fig:Computed-linear-optical_ppp-CI_std}, we present our calculated spectra using the screened
and standard parameters, respectively, within the PPP-CI approach.
If we compare the relative intensity of the first peak of the experimental
spectra, we find that results of Arabei $et$ $al$.\cite{arabei2000}
are in perfect agreement with our results in that the first peak is
not the most intense. However, Konishi $et$ $al.$\cite{Konishi_Bisanthenes}
report that the first peak is the most intense one, in complete disagreement
with our results. Our calculated location of the first peak corresponding
to the optical gap, was found to be $2.00$ eV with the screened parameters,
and $2.21$ eV for the standard parameters. As is obvious from Table
\ref{tab:Comparison-RGM}, the experimental values of the optical
gap range from $1.80$ eV to $2.15$ eV. Thus, we find that both our
screened and standard parameter of optical gap are quite close to
the range of experimental values. We also note that our screened parameter
of $2.00$ eV is in almost perfect agreement with the value of optical
gap $2.02$ eV, measured by Clar and Schmidt\cite{clar1977corr}.
{As far as higher energy peaks are concerned, Konishi
$et$ $al.$\cite{Konishi_Bisanthenes} report a peak at 4.05 eV,
which is in good agreement with our standard parameter peak computed
at 4.14 eV, while the corresponding screened parameter candidate at
4.19 eV is somewhat higher. Our calculation predicts several more
peaks, whose details are given in Table S4 of Supporting Information.}
{We note that some of these peaks are in an energy
range, for which no experimental results exist. }We hope that in future
measurements of the absorption spectrum of bisanthenes, energy range
of 5 eV, and beyond, will be explored. 

Peak VI is the most intense peak in the absorption spectra computed
using both the screened as well as the standard parameters. The most
intense peak computed using the screened parameters located at 5.63
eV, corresponds to a state with $B_{3u}$ symmetry, whose wave function
is dominated by $\left|H-2\rightarrow L+4\right\rangle $$-c.c.$
excitations, where $c.c.$ denotes the charge conjugated configration.
However, the standard parameter calculations predict the most intense
intense peak to be due to a $B_{3u}$ state, located at 5.95 eV, along
with a small mixture of a $B_{2u}$ state located at 6.17 eV, with
their wave functions dominated by single excitations $\left|H-2\rightarrow L+4\right\rangle $$-c.c.$,
and $\left|H-3\rightarrow L+3\right\rangle $, respectively. The detailed
wave function analysis of all the excited states contributing to various
peaks in the calculated spectra of RGM-28, is presented in Tables
S12 and S13 of the Supporting Information.

\subsubsection*{RGM-30}

Koch $et$ $al.$\cite{koch1991polyarylenes}, Ruiterkamp $et$ $al.$\cite{ruiterkamp2002spectroscopy}
and Halasinski $et$ $al.$\cite{halasinski2003electronic} have
reported the measurements of the absorption spectrum of terrylene,
the structural analog of RGM-$30$, and its derivatives. However,
Clar $et$ $al.$\cite{clar1978correlations}, Kummer $et$ $al.$\cite{kummer1997absorption},
Biktchantaev $et$ $al.$\cite{biktchantaev2002perylene} and Baumgarten
$et$ $al.$\cite{baumgarten1994spin} reported only the optical
gap of terrylene. In Figs. \ref{fig:Computed-linear-optical_ppp-CI_scr}
 and \ref{fig:Computed-linear-optical_ppp-CI_std}, we present our calculated spectra using the screened
and standard parameters, respectively, within the PPP-CI approach.
If we compare the relative intensity of the first peak of the experimental
spectra, we find that the results of Koch $et$ $al.$\cite{koch1991polyarylenes},
Ruiterkamp $et$ $al.$\cite{ruiterkamp2002spectroscopy} and Halasinski
$et$ $al.$\cite{halasinski2003electronic} are in perfect agreement
with our results in that the first peak is not the most intense. The
calculated location of the first peak of the absorption spectrum,
which defines the optical gap, was found to be $2.11$ eV, and $2.43$
eV, from our standard, and screened parameter based calculations,
respectively. As it is obvious from Table \ref{tab:Comparison-RGM},
that the experimental values of the optical gap range from $2.14$
eV to $2.36$ eV. Thus, we find that both our screened and standard
parameter of optical gap are quite close to the range of experimental
values. We also note that our screened parameter value of $2.11$
eV is in almost perfect agreement with the value of optical gap $2.14$
eV, measured by Kummer $et$ $al.$\cite{kummer1997absorption}.
{As far as higher energy peaks are concerned, Halasinski
$et$ $al.$\cite{halasinski2003electronic} report a peak at 5.41
eV, in good agreement with our screened parameter peak computed at
5.35 eV, while the corresponding standard parameter candidate at 5.53
eV is in good agreement with a peak at 5.48 eV, measured by Koch $et$
$al.$\cite{koch1991polyarylenes}. Our calculation predicts several
more peaks, whose details are given in Table S5 of Supporting Information.}
Furthermore, we have computed several peaks located beyond 7 eV, for
which no experimental results exist. We hope that in future measurements
of the absorption spectrum of terrylene, this higher energy range
will be explored. 

In the spectra computed using the screened parameters, IV peak is
the most intense, and it is due to a $B_{2u}$ state, located at 5.12
eV, whose wave function is dominated by the $\left|H-2\rightarrow L+2\right\rangle $
excitation. For the standard parameter calculations, peak V is the
most intense one, due to a $B_{2u}$ state located at 6.01 eV, along
with a small mixture of $B_{3u}$ state located at 5.87 eV, with wave
functions dominated by configurations $\left|H-3\rightarrow L+3\right\rangle $,
and $\left|H\text{\ensuremath{\rightarrow}}L;H\rightarrow L+4\right\rangle $$-c.c.$,
respectively. The detailed wave analysis of all the excited states
contributing peaks in the computed spectra is presented in Tables
S14 and S15 of the Supporting Information. 

\subsubsection*{RGM-36}

The hydrogen passivated structural analog of RGM-36 is tetrabenzocoronene,
for which we were unable to locate any experimentally measured optical
absorption spectrum. Therefore, we can only compare our calculations
to the theoretical works of other authors, for which also we could
find just one TDDFT based computation of the optical absorption spectra
by Malloci $et$ $al.$\cite{online_RQD36}. In Figs. \ref{fig:Computed-linear-optical_ppp-CI_scr}
 and \ref{fig:Computed-linear-optical_ppp-CI_std}, we present our calculated spectra using the screened
and standard parameters, respectively, within the PPP-CI approach.
If we compare the relative intensity of the first peak in the spectra,
we find that the results of Malloci $et$ $al.$\cite{online_RQD36}
are in perfect agreement with our results in that the first peak is
not the most intense. In Table \ref{tab:Comparison-RGM}, we have
compared the locations of various peaks reported by Malloci $et$
$al.$ with our computed results. We find that the value of optical
gap reported by Malloci $et$ $al.$ \cite{malloci2011electronic}
is 0.95 eV, which is significantly smaller than our computed results
of 2.11 eV (screened) and 2.30 eV (standard). Given such severe disagreement
between two theoretical calculations, it will be really useful if
an experiment is performed on this molecule, or another theoretical
calculation is done. Given the fact that our results on optical gaps
on smaller RGMs were in excellent agreement with the experiments,
we speculate that the TDDFT calculation of Malloci $et$ $al.$\cite{online_RQD36}
has significantly underestimated the optical gap of RGM-36.{{}
As far as higher peaks are concerned, our screened parameter calculations
predict a peak at 3.63 eV due to a $^{1}B_{3u}$ state, whose location
is in perfect agreement with a peak at 3.64 eV, reported by Malloci
$et$ $al.$ \cite{online_RQD36}. Our calculation predicts several
more peaks, whose details are given in Table S6 of Supporting Information.}
As far as higher energy peaks computed by Malloci $et$ $al.$\cite{online_RQD36}
are concerned, our PPP model values are generally in good agreement
with them. 

For both the screened as well as the standard parameters computed
absorption spectra, peak VIII is the most intense one. In the spectra
computed using the screened parameters, the most intense peak is located
at 5.99 eV, and is due to states of symmetries $B_{2u}$ and $B_{3u}$
contributing almost equally to the oscillator strength, with their
many-particle wave functions dominated by configurations $\left|H-3\rightarrow L+3\right\rangle $,
and $\left|H-2\rightarrow L+5\right\rangle $$+c.c.$, respectively.
In the spectrum computed using the standard parameters, the most intense
peak is located at 6.57 eV, due to a $B_{3u}$ state, with wave function
dominated by $\left|H-5\rightarrow L+2\right\rangle $$+c.c.$ excitations.
The detailed wave analysis of excited states contributing peaks in
the spectra computed by the screened, and the standard parameters
is presented in Tables S16 and S17 of the Supporting Information. 

\subsubsection*{RGM-40}

Ruiterkamp $et$ $al.$\cite{ruiterkamp2002spectroscopy}, Koch $et$
$al.$\cite{koch1991polyarylenes}, and Halasinski $et$ $al.$ \cite{halasinski2003electronic}
have reported the measurements of the absorption spectrum of quaterrylene,
the hydrogen passivated structural analog of RGM-$40$, and its derivatives.
However, Clar $et$ $al.$\cite{clar1978correlations}, Former $et$
$al.$ \cite{former2002cyclodehydrogenation}, Gudipati $et$ $al.$
\cite{gudipati2006double}, and Baumgarten $et$ $al.$\cite{baumgarten1994spin}
reported only the optical gap of quaterrylene. In Figs. \ref{fig:Computed-linear-optical_ppp-CI_scr}
 and \ref{fig:Computed-linear-optical_ppp-CI_std}, we present our calculated spectra using the screened
and the standard parameters, respectively, within the PPP-CI approach.
If we compare the relative intensity of the first peak of the experimental
spectra, we find that results of Ruiterkamp $et$ $al.$\cite{ruiterkamp2002spectroscopy}
and Halasinski $et$ $al.$ \cite{halasinski2003electronic} are
in perfect agreement with our results in that the first peak is not
the most intense. However, Koch $et$ $al.$\cite{koch1991polyarylenes}
report that the first peak is the most intense one, in disagreement
with our results, as well those of other experimentalists. As is obvious
from Table \ref{tab:Comparison-RGM}, the experimental values of the
optical gap range from $1.87$ eV to $2.04$ eV, implying that our
screened parameter result of optical gap (2.02 eV) is quite close
to the range of experimental values, while the optical gap obtained
using the standard parameters (2.30 eV) is somewhat larger. We also
note that our screened parameter value of the optical gap, $2.02$
eV, is in almost perfect agreement with 2.03 eV, the value of the
optical gap measured by Clar and Schmidt\cite{clar1977corr}. {As
far as the higher energy peaks are concerned, Koch $et$ $al$.\cite{koch1991polyarylenes}
report a peak at 5.27 eV, in good agreement with our screened-parameter
based peak at 5.16 eV. Halasinski $et$ $al.$ \cite{halasinski2003electronic}
report a peak at 5.39 eV, which is in good agreement with the peak
at 5.48 eV, predicted by standard parameter calculations. Our calculation
predicts several more peaks, whose details are given in Table S7 of
Supporting Information.} Furthermore, Halasinski $et$ $al.$ \cite{halasinski2003electronic}
have measured four more peaks in the range 5.82 --- 6.63 eV, each
of which is in good agreement with our calculated peaks (see Table
S7 of Supporting Information). 

In the spectra computed using the screened parameters, peak IV is
most intense, and it is due to a state of $B_{2u}$ symmetry, located
at 5.16 eV, along with a small contribution from a $B_{3u}$ state,
located at 5.17 eV. The wave functions of these states are dominated
by single excitations $\left|H-2\rightarrow L+2\right\rangle $, and
$\left|H-1\text{\ensuremath{\rightarrow}}L+5\right\rangle $ $-c.c.$,
respectively. For the standard parameter calculations, peak VI is
the most intense one, corresponding again to a mixture of a $B_{2u}$
state (at 6.23 eV), and a $B_{3u}$ state (at 6.17 eV), with wave
functions dominated by excitations $\left|H-4\rightarrow L+4\right\rangle ,$
and $\left|H\rightarrow L+1;H-6\rightarrow L\right\rangle $ $-c.c.$,
respectively. The detailed wave function analysis of the excited states
contributing to various peaks in the spectra calculated by the screened
and standard parameters, respectively, is presented in Tables S18
and S19 of the Supporting Information.

\subsubsection*{RGM-42}

Teranthene is the hydrogen-saturated structural analogue of RGM-42,
for which no experimental, or theoretical data is available, as far
as optical absorption spectrum is concerned. However, Konishi $et$
$al.$ \cite{konishi2014anthenes} have reported the measurement
of the absorption spectrum of teranthene with tertiary-butyl group
attached on its edge atoms, and the results of their experiments,
along with those obtained from our calculations, are summarized in
Table \ref{tab:Comparison-RGM}. In Figs. \ref{fig:Computed-linear-optical_ppp-CI_scr}
 and \ref{fig:Computed-linear-optical_ppp-CI_std}, we present our calculated spectra of RGM-42 using the
screened, and the standard parameters, respectively, within the PPP-CI
approach. If we compare the relative intensity of the first peak of
the experimental spectra, we find that results of Konishi $et$ $al.$
\cite{konishi2014anthenes} are in perfect agreement with our results
in that the first peak is not the most intense. However, as is obvious
from Table \ref{tab:Comparison-RGM}, quantitatively speaking, our
theoretical results and experimental results of Konishi $et$ $al.$\cite{konishi2014anthenes}
disagree completely in the low-energy region. Konishi $et$ $al.$\cite{konishi2014anthenes}
have reported two low-lying excited energy peaks located at 1.17 eV,
and 1.21 eV, for which there are no counterparts in our computed spectra.
The calculated location of the first peak, which also corresponds
to the optical gap, was found to be $1.86$ eV from our screened parameter
calculation, and $2.04$ eV for the standard parameter calculation,
as against significantly smaller values 1.17--1.21 eV measured by
Konishi $et$ $al.$ \cite{konishi2014anthenes}. {As
far as higher energy peaks are concerned, Konishi $et$ $al.$ \cite{konishi2014anthenes}
report a peak at 3.87 eV, which is in good agreement both with a screened
parameter peak at 3.96 eV, and a standard parameter peak at 3.80 eV.
Our calculations predict several peaks in the energy region of 4 eV
and beyond (see Table S8 of Supporting Information for details), which
Konishi }{\emph{et al.}}{\cite{konishi2014anthenes}
have not probed.} The only possible reason we can think of behind
the disagreement between the theory and the experiments in the lower
energy region is that the experiments were performed on teranthene
saturated with t-butyl group, while our results are valid for hydrogen-saturated
material. Nevertheless, we hope that more groups will perform measurements
of optical absorption spectra of teranthenes so as to be sure about
the value of their optical gap.

In the spectrum computed using the screened parameters, peak III is
most intense, corresponding to a $B_{3u}$ state located at 3.96,
whose wave function is dominated by $\left|H-3\text{\ensuremath{\rightarrow}}L\right\rangle $
$-c.c.$ single excitations. For the standard parameter calculation,
peak VII is most intense, corresponding again to a $B_{3u}$ state,
but located at 5.47 eV, with a small mixture of $B_{2u}$ state located
at 5.33 eV. The wave functions of the two states are dominated by
single excitations, $\left|H-3\rightarrow L\right\rangle $$-c.c.$,
and $\left|H-1\rightarrow L+1\right\rangle $, respectively. The detailed
wave analysis of all the excited states contributing to peaks in the
calculated spectra using the screened and the standard paremeters,
is presented in Tables S20 and S21, respectively, of the Supporting Information.. 

\subsubsection*{RGM-50}

For RGM-50 and larger structures, it would have been computationally
very tedious to perform MRSDCI calculations retaining all the MOs,
therefore, we decided to freeze a few lowest-lying occupied orbitals,
and delete their electron-hole symmetric highest virtual orbitals.
For the case of RGM-50, we froze/deleted four occupied/virtual orbitals,
so that the total number of MOs involved in the calculations reduced
to forty-two, same as in case of RGM-42. In order to demonstrate that
this act of freezing and deleting the MOs does not affect the calculated
optical absorption spectra, we have performed the calculations for
RGM-50, with four and seven frozen/deleted orbitals, leading to 42/36
active MOs. {{} From the calculated absorption spectra presented in Fig.
\ref{fig:Convergence-of-the-spectra}, it is obvious that except for
the intensity of the highest energy peak IX in the standard parameter
calculations, the spectra remain the same for both the cases, implying
that the convergence has been achieved within an acceptable tolerance}.

\begin{figure}
\includegraphics[scale=0.6]{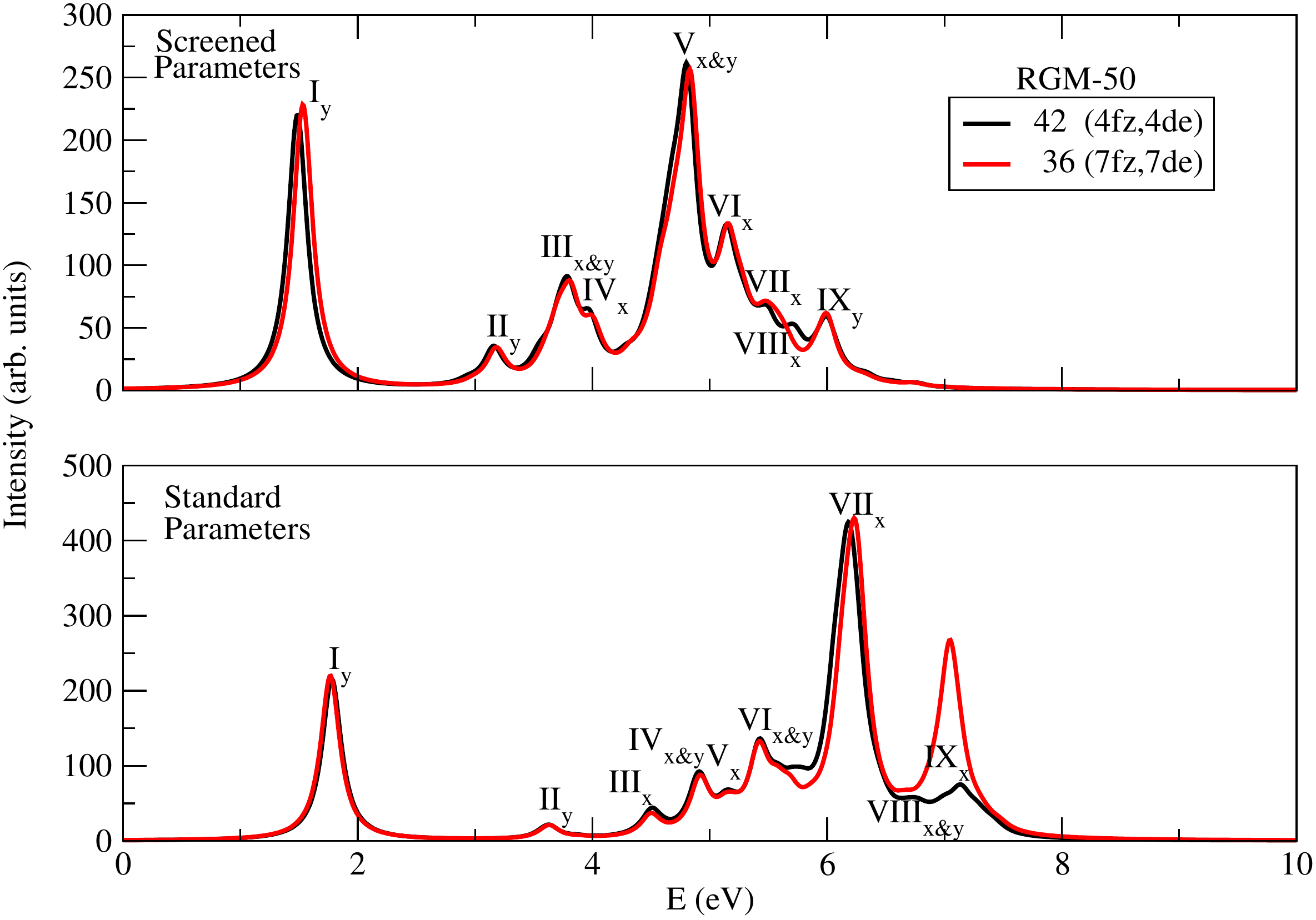}

\caption{Convergence of the calculated spectra of RGM-50, with respect to the
number of active orbitals in the MRSDCI calculations. Curve in black
color corresponds to the calculation with 42 active orbitals (4 frozen
and 4 deleted orbitals), while that in red is from a calculation
with 36 active orbitals (7 frozen and 7 deleted orbitals).\label{fig:Convergence-of-the-spectra}}
\end{figure}
Hydrogen saturated structural analog of RGM-50 is pentarylene, for
which, to the best of our knowledge, no experimental data of optical
absorption exists. However, Koch $et$ $al.$\cite{koch1991polyarylenes}
have reported the measurements of the absorption spectrum of pentarylene
saturated with the t-butyl group, while Baumgarten $et$ $al.$\cite{baumgarten1994spin}
measured only its optical gap. We present our calculated absorption
spectra for RGM-50 in Figs. \ref{fig:Computed-linear-optical_ppp-CI_scr}
 and \ref{fig:Computed-linear-optical_ppp-CI_std}, while the comparison of important peak locations resulting
from our calculations, with the experiments, and other theoretical
works is presented in Table \ref{tab:Comparison-RGM}. The value of
the optical gap measured by both the groups\cite{koch1991polyarylenes,baumgarten1994spin}
is 1.66 eV, which is in very good agreement with the value 1.72 eV
computed using the screened parameters, while the corresponding standard
parameter value 1.98 eV is on the higher side. If we compare the relative
intensity of the first peak of the experimental spectra corresponding
to the optical gap, we find that Koch $et$ $al.$\cite{koch1991polyarylenes}
report that the first peak to be the most intense one, in disagreement
with our results. However, the noteworthy point is that the first
peak computed using the screened parameter, is quite intense, and
is only somewhat lesser in intensity than the most intense peak (peak
V) of the computed spectrum. {As far as higher energy
peaks are concerned, Koch $et$ $al$.\cite{koch1991polyarylenes}
report a peak at 4.80 eV, which is somewhat close to our screened
parameter peak located at 4.97 eV. } Furthermore, we have computed
several higher energy peaks as well, for which no experimental results
exist. We hope that in future measurements of the absorption spectrum
of pentarylene, energy range beyond 5.3 eV will be explored. 

On comparing our results to the calculations by other authors, we
find that the value of the optical gaps reported by Viruela-Mart\'{i}n
$et$ $al.$ \cite{viruela1992theoretical} (1.40 eV) using the valence-effective
Hamiltonian approach, Malloci $et$ $al.$ \cite{malloci2011electronic}
(1.54 eV) using the TDDFT method are significantly smaller than our
results, as well as experiments. However, Minami $et$ $al.$ \cite{minami2012theoretical}
report a TDDFT value which is in good agreement with the experiment
value of the optical gap, but about 0.1 eV lower than our result.
Karabunarliev $et$ $al.$\cite{karabunarliev1994structure} computed
the optical gap to be 1.97 eV using PM3 semi-empirical method, is
in perfect agreement with our standard parameter result located at
1.98 eV, but significantly higher than the experimental value, as
well as our screened parameter value. As far as higher energy peaks
computed by Malloci $et$ $al.$ are concerned, our PPP model values
are in reasonable agreement with them. 

In the spectra computed using the screened parameter, peak V is most
intense, and is due to a $B_{2u}$ state located at 5.04 eV, along
with a small intensity due to a $B_{3u}$ state located at 4.90 eV.
The wave functions of the two states are dominated by configurations
$\left|H-3\rightarrow L+3\right\rangle $, and $\left|H-5\rightarrow L+1\right\rangle $$+c.c.$
excitations, respectively. In the standard parameter spectrum, peak
VIII is most intense, and is mainly due to a $B_{2u}$ state located
at 6.38 eV, along with a small contribution of a $B_{3u}$ state located
at 6.45 eV. The dominant contributions to the many-particle wave functions
of these two states are from single excitation $\left|H-4\rightarrow L+4\right\rangle $,
and the double excitation $\left|H\rightarrow L+6;H\rightarrow L+1\right\rangle $$-c.c.$,
respectively. The detailed wave function analysis of all the excited
states contributing peaks in the spectra computed using the screened
and the standard parameters, are presented in Tables S22, and S23,
respectively, of the Supporting Information.

\subsubsection*{RGM-54 }

For the case of RGM-54, we performed MRSDCI calculations after freezing/deleting
six occupied/virtual MOs, i.e., with forty two active MOs. For this
molecule, we were not able to locate any experimental results, or
other theoretical calculations, thus, making our results the first
ones. We hope that our calculations will give rise to future theoretical
and experimental works on this system.

Calculated optical absorption spectra for RGM-54 are presented in
Figs. \ref{fig:Computed-linear-optical_ppp-CI_scr}  and \ref{fig:Computed-linear-optical_ppp-CI_std}, obtained
using the screened and standard parameters, respectively. {The
locations of important peaks, and the symmetries of excited states
giving rise to them, are summarized in Table S10 of Supporting Information}.
The first peak corresponding to the optical gap{{} (see
Table \ref{tab:Comparison-RGM})}, is a very weak peak from both sets
of calculations, and was found to be at 1.63 eV with the screened
parameters, and 2.09 eV with the standard parameters. Given the pattern
observed for smaller RGMs discussed in the previous sections, we expect
the screened parameter value of the optical gap to be closer to the
experimental value. 

In the screened parameter calculations, next we find a group of three
well-separated peaks, with strong, and almost equal, intensities,
located at 2.56 eV, 2.83 eV, and 3.09 eV. The first of
these peaks corresponds to an $x$-polarized transition, while the
next two are $y$-polarized. In the standard parameter spectrum as
well, the next three peaks are quite strong, and well separated, but
they have their intensities in the ascending order, while the middle
peak (peak III) appears as a shoulder of peak IV. The locations of
these peaks are blue-shifted compared to their screened parameter
counterparts, and are 3.20 eV, 3.69 eV, and 3.98 eV. The polarization
characteristics are also different, with two of the peaks exhibiting
mixed polarization.

At higher energies, in the screened parameter spectrum there are well
defined high-intensity peaks at energies 3.71 eV ($x/y$ polarized),
3.95 eV ($x$ polarized), 5.14 eV ($y$ polarized), and 5.40 eV ($x$
polarized). Out of these the peak located at 5.14 eV (peak IX) is
the most intense peak of the computed spectrum. This peak is due to
a $B_{2u}$ state, whose wave function is dominated by the $\left|H-3\rightarrow L+3\right\rangle $
singly-excited configuration.

In the standard parameter spectrum, beyond 4 eV, there are a number
of low-intensity peaks or shoulders, except for a peak located at
6.56 eV (peak XII), which is the most intense one, and exhibits mixed
polarization. This peak is due to a $B_{3u}$ state located at 6.51
eV, along with a smaller contribution to the intensity from a $B_{2u}$
state located at 6.61 eV. The wave functions of the two states are
dominated by singly excited configurations, $\left|H-2\rightarrow L+7\right\rangle $$+c.c.$,
and $\left|H-3\rightarrow L+3\right\rangle $, respectively. 

The detailed wave function analysis of all the excited states contributing
peaks in the spectra computed using the standard parameters, and the
screened parameters, are presented in Tables S24, and S25, respectively,
of the Supporting Information.

\subsubsection*{RGM-56}

Again, due to a large number of electrons in the system, for RGM-56
we froze/deleted seven occupied/virtual orbitals, so that the total
number of active MOs involved in the calculations reduced to forty-two,
same as in case of RGM-42, RGM-50, and RGM-54. 

The hydrogen saturated analog of RGM-56 is quateranthene, for which
no experimental measurements of optical absorption spectrum exist.
However, Konishi $et$ $al.$\cite{konishi2014anthenes} measured
the absorption spectrum of quateranthene, with t-butyl groups attached
to its edge carbon atoms, with which we will compare our calculated
spectra. In Figs. \ref{fig:Computed-linear-optical_ppp-CI_scr}  and
\ref{fig:Computed-linear-optical_ppp-CI_std}, we present our calculated spectra using the screened and standard
parameters, respectively, within the PPP-MRSDCI approach. In Table
\ref{tab:Comparison-RGM}, we present the locations of various peaks
in the calculated spectra, and compare them to the measured values
of Konishi $et$ $al.$\cite{konishi2014anthenes}. If we compare
the relative intensity of the first peak of the experimental spectra,
we find that the results of Konishi $et$ $al.$\cite{konishi2014anthenes}
are in perfect agreement with ours in that the first peak is not the
most intense. The calculated location of the first peak, which also
corresponds to the optical gap, from our calculations was found to
be 1.50 eV with the screened parameters, and 1.91 eV with the standard
parameters. The experimental value of optical gap reported by Konishi
$et$ $al.$\cite{konishi2014anthenes} is 1.35 eV, which is about
0.15 eV lower than our screened parameter value, but significantly
smaller than the value obtained from the standard parameter calculations.{{}
As far as higher energy peaks are concerned, Konishi $et$ $al.$\cite{konishi2014anthenes}
report a peak at 3.21 eV, in good agreement with our standard parameter
peak computed at 3.35 eV.. Our calculation predicts several more peaks,
whose details are given in Table S11 of Supporting Information.}  We
hope that in future measurements of the absorption spectrum of quateranthene,
energy range beyond 3.50 eV will be explored. 

In the spectra computed using screened parameter, peak III is the
most intense one, and it is due to a $B_{2u}$ state located at 2.76
eV, with a smaller contribution to the intensity from a $B_{3u}$
state located at 2.82 eV. The wave functions of the two states are
dominated by doubly-excited configurations $\left|H\rightarrow L;H\rightarrow L+1\right\rangle $$-c.c.$,
and $\left|H\rightarrow L;H-2\rightarrow L\right\rangle $$-c.c.$,
respectively. In the standard parameter spectrum, peak V is most intense,
and is entirely due to a $B_{2u}$ state located at 4.34 eV, whose
wave function derives most important contribution from the singly-excited
configuration $\left|H-2\rightarrow L+2\right\rangle $. The detailed
wave function analysis of the excited states contributing to varioius
peaks in the spectra computed using the screened and standard parameters
is presented in Tables S26, and S27, respectively, of the Supporting
Information.

\section*{Conclusions}

\label{sec:conclusions}

In this paper, we have presented the results of our  correlated-electron
calculations of spin gaps and optical absorption spectra of rectangular
graphene-like polycyclic aromatic molecules, with the number of carbon
atoms in the range 28--56, using PPP model Hamiltonian, and the MRSDCI
approach. We analyzed the ground state wave functions of these molecules,
and found that with the increasing size, they exhibit significant
configuration mixing leading to diradical open-shell character. Results
of our calculations on the spin gaps of these RGMs, when extrapolated
to infinite graphene, suggest that it has a vanishing spin gap, implying
weak electron correlation effects. This result is consistent with
the widespread assumption that graphene is a weakly-correlated material.

For the case of optical absorption spectra, we generally found very
good agreement with the experiments performed on hydrogen-saturated
structural analog of each RGM, wherever experimental data was available.
In certain cases, where no experimental data was available for the
H-passivated molecule, the comparison was instead made with the measurements
performed on t-butyl group saturated systems, and some quantitative
disagreements were encountered, most severe of which were for RGM-42.
It will be very interesting if future experimental measurements could
be performed on the H-passivated molecules in those cases. For the
case of RGM-36, and RGM-54, no experimental measurements exist, while
for the case of RGM-54, even prior theoretical calculations do not
exist. Thus, results of our calculations on these molecules could
be tested in  future measurements of their absorption spectra.

\bibliography{rgm}

\section*{Acknowledgements}

This research was supported in part by a grant from SERB-DST, Government
of India, under project no. SB/S2/CMP-066/2013. 

\section*{Author Contributions Statement}

A.S. conceived of the problem and D.K.R. performed the necessary calculations.
Both the authors contributed to the analysis of the results and writing
of the manuscript.

\section*{Additional Information}

Most of the data associated with the calculations has been made available
in the accompanying Supporting Information. Any other data will be
made available on demand. \\* {\textbf{Competing Interests:}} The authors declare no competing interests.


\end{document}


\flushbottom
\maketitle
%
%
\section*{Size of the CI matrices}

In order to explain the large-scale nature of these calculations,
in Tables \ref{size_of_matrix-for_spin_gapRGQD} and \ref{size_of_matrix-RGQD}
and we present the dimensions of the CI matrices employed in these
calculations, for various symmetry subspaces, of different RGMs. The
CI calculations are performed using MRSDCI method, by employing PPP
Coulomb parameters, and the point group symmetry of the concerned
RGMs, are also indicated in the table. 

\begin{table}[H]
\caption{\label{size_of_matrix-for_spin_gapRGQD}Dimensions of the CI matrices
($N_{total}$) for different symmetry subspaces, employed in the calculations
of spin gaps of RGMs containing $n$ carbon atoms.}

\centering{}%
\begin{tabular}{|c|c|c|}
\hline 
\multirow{2}{*}{$n$} & \multicolumn{1}{c|}{$1^{1}A_{g}$ } & \multicolumn{1}{c|}{$1^{3}B_{2u}$}\tabularnewline
\cline{2-3} 
 & $N_{total}$ & $N_{total}$\tabularnewline
\hline 
\multirow{2}{*}{$28$} & $1630819^{a}$ & $3363548^{a}$\tabularnewline
\cline{2-3} 
 & $929383^{b}$ & $2052250^{b}$\tabularnewline
\hline 
\multirow{2}{*}{$30$} & $1271636^{a}$ & $2516476^{a}$\tabularnewline
\cline{2-3} 
 & $1355882^{b}$ & $2224033^{b}$\tabularnewline
\hline 
\multirow{2}{*}{$36$} & $2895688^{a}$ & $6999314^{a}$\tabularnewline
\cline{2-3} 
 & $3470683^{b}$ & $8213570^{b}$\tabularnewline
\hline 
\multirow{2}{*}{$40$} & $5346813^{a}$ & $9103631^{a}$\tabularnewline
\cline{2-3} 
 & $3897268^{b}$ & $4429759^{b}$\tabularnewline
\hline 
\multirow{2}{*}{$42$} & $4690547^{a}$ & $8964411^{a}$\tabularnewline
\cline{2-3} 
 & $5694171^{b}$ & $10937840^{b}$\tabularnewline
\hline 
\multirow{2}{*}{$50$} & $3504598^{a}$ & $3371187^{a}$\tabularnewline
\cline{2-3} 
 & $2655705^{b}$ & $5103671^{b}$\tabularnewline
\hline 
\multirow{2}{*}{$54$} & $4479514^{a}$ & $10745638^{a}$\tabularnewline
\cline{2-3} 
 & $3949223^{b}$ & $9923969^{b}$\tabularnewline
\hline 
\multirow{2}{*}{$56$} & $3913511^{a}$ & $10766994^{a}$\tabularnewline
\cline{2-3} 
 & $4181480^{b}$ & $9476458^{b}$\tabularnewline
\hline 
\multirow{2}{*}{72} & $4180503^{a}$ & $10475673^{a}$\tabularnewline
\cline{2-3} 
 & $4171695^{b}$ & $9450095^{b}$\tabularnewline
\hline 
\multicolumn{3}{|c|}{MRSDCI$^{a}$ method with screened parameters. }\tabularnewline
\multicolumn{3}{|c|}{MRSDCI$^{b}$ method with standard parameters.}\tabularnewline
\hline 
\end{tabular}\protect
\end{table}
\begin{table}[H]
\caption{\label{size_of_matrix-RGQD} Dimension of the CI matrices ($N_{total}$)
of different symmetry subspaces involved in the MRSDCI calculations
of the optical absorption spectra of RGMs, containing $n$ carbon
atoms.}

\centering{}%
\begin{tabular}{|c|c|c|c|}
\hline 
\multirow{2}{*}{n} & \multicolumn{1}{c|}{$^{1}A_{g}$ } & \multicolumn{1}{c|}{$^{1}B_{2u}$ } & \multicolumn{1}{c|}{$^{1}B_{3u}$ }\tabularnewline
\cline{2-4} 
 & $N_{total}$ & $N_{total}$ & $N_{total}$\tabularnewline
\hline 
\multirow{2}{*}{$28$} & $243473^{a}$ & $1286950^{a}$ & $1195300^{a}$\tabularnewline
\cline{2-4} 
 & $176670^{b}$ & $1304940^{b}$ & $1363812^{b}$\tabularnewline
\hline 
\multirow{2}{*}{$30$} & $148576^{a}$ & $1417769^{a}$ & $1128568^{a}$\tabularnewline
\cline{2-4} 
 & $210078^{b}$ & $1650565^{b}$ & $1220974^{b}$\tabularnewline
\hline 
\multirow{2}{*}{$36$} & $1093085^{a}$ & $1309198^{a}$ & $1506432^{a}$\tabularnewline
\cline{2-4} 
 & $1149578^{b}$ & $1983952^{b}$ & $1607780^{b}$\tabularnewline
\hline 
\multirow{2}{*}{$40$} & $2675017^{a}$ & $4768452^{a}$ & $3136672^{a}$\tabularnewline
\cline{2-4} 
 & $3897268^{b}$ & $6200883^{b}$ & $4820202^{b}$\tabularnewline
\hline 
\multirow{2}{*}{$42$} & $2729332^{a}$ & $3551497^{a}$ & $3195172^{a}$\tabularnewline
\cline{2-4} 
 & $3477479^{b}$ & $395229^{b}$ & $4042750^{b}$\tabularnewline
\hline 
\multirow{2}{*}{$50$} & $3504598^{a}$ & $4494916^{a}$ & $5083760^{a}$\tabularnewline
\cline{2-4} 
 & $2655705^{b}$ & $5637511^{b}$ & $6193064^{b}$\tabularnewline
\hline 
\multirow{2}{*}{$54$} & $2150043^{a}$ & $2473193^{a}$ & $3038570^{a}$\tabularnewline
\cline{2-4} 
 & $3047218^{b}$ & $4003216^{b}$ & $357911^{b}$\tabularnewline
\hline 
\multirow{2}{*}{$56$} & $2125583^{a}$ & $2983416^{a}$ & $3013058^{a}$\tabularnewline
\cline{2-4} 
 & $2830899^{b}$ & $3086912^{b}$ & $3308186^{b}$\tabularnewline
\hline 
\multicolumn{4}{|c|}{MRSDCI$^{a}$ method with screened parameters. }\tabularnewline
\multicolumn{4}{|c|}{MRSDCI$^{b}$ method with standard parameters.}\tabularnewline
\hline 
\end{tabular}\protect
\end{table}

\section*{Orbital Occupation}

\begin{table}[H]
\centering{}\caption{The table presents the number of reference configurations ($N_{ref}$)
used for the MRSDCI calculations for the ground state of various RGMs,
performed using both the standard and the screened parameters in the
PPP model. Additionally, the total number of doubly and singly occupied
orbitals in the reference configurations, classified according to
their irreducible representations (irreps) of the $D_{2h}$ point
group, are also presented for each calculation. Those configurations
are chosen as reference configurations, if the magnitude of their
coefficient in the ground state wave functions is at least 0.05.}
\begin{tabular}{|c|c|c|c|c|c|c|c|c|c|c|c|c|}
\hline 
\multirow{3}{*}{System} & \multicolumn{6}{c|}{Screened parameters} & \multicolumn{6}{c|}{Standard parameters}\tabularnewline
\cline{2-13} 
 & $N_{ref}$ & Orbital & \multicolumn{4}{c|}{Orbital Irreps} & $N_{ref}$ & Orbital & \multicolumn{4}{c|}{Orbital Irreps}\tabularnewline
\cline{4-7} \cline{10-13} 
 &  & Occupancy & $^{1}A_{g}$ & $^{1}B_{3u}$ & $^{1}B_{2u}$ & $^{1}B_{1g}$ &  & Occupancy & $^{1}A_{g}$ & $^{1}B_{3u}$ & $^{1}B_{2u}$ & $^{1}B_{1g}$\tabularnewline
\hline 
\multirow{2}{*}{RGM-28} & \multirow{2}{*}{12} & Double & 3 & 2 & 2 & 2 & \multirow{2}{*}{9} & Double & 3 & 2 & 2 & 2\tabularnewline
\cline{3-7} \cline{9-13} 
 &  & Single & 1 & 1 & 2 & 1 &  & Single & 1 & 1 & 2 & 1\tabularnewline
\hline 
\multirow{2}{*}{RGM-30} & \multirow{2}{*}{6} & Double & 4 & 2 & 3 & 1 & \multirow{2}{*}{8} & Double & 4 & 2 & 3 & 1\tabularnewline
\cline{3-7} \cline{9-13} 
 &  & Single & 1 & 1 & 1 & 2 &  & Single & 1 & 1 & 1 & 2\tabularnewline
\hline 
\multirow{2}{*}{RGM-36} & \multirow{2}{*}{4} & Double & 5 & 3 & 4 & 3 & \multirow{2}{*}{6} & Double & 4 & 3 & 4 & 2\tabularnewline
\cline{3-7} \cline{9-13} 
 &  & Single & 0 & 1 & 1 & 1 &  & Single & 1 & 1 & 1 & 2\tabularnewline
\hline 
\multirow{2}{*}{RGM-40} & \multirow{2}{*}{5} & Double & 6 & 3 & 5 & 2 & \multirow{2}{*}{7} & Double & 5 & 3 & 5 & 2\tabularnewline
\cline{3-7} \cline{9-13} 
 &  & Single & 0 & 1 & 1 & 2 &  & Single & 1 & 1 & 1 & 2\tabularnewline
\hline 
\multirow{2}{*}{RGM-42} & \multirow{2}{*}{4} & Double & 5 & 5 & 5 & 3 & \multirow{2}{*}{6} & Double & 5 & 4 & 4 & 3\tabularnewline
\cline{3-7} \cline{9-13} 
 &  & Single & 1 & 0 & 1 & 1 &  & Single & 1 & 1 & 2 & 1\tabularnewline
\hline 
\multirow{2}{*}{RGM-50} & \multirow{2}{*}{4} & Double & 8 & 4 & 7 & 3 & \multirow{2}{*}{4} & Double & 8 & 4 & 7 & 3\tabularnewline
\cline{3-7} \cline{9-13} 
 &  & Single & 0 & 1 & 0 & 2 &  & Single & 0 & 1 & 0 & 2\tabularnewline
\hline 
\multirow{2}{*}{RGM-54} & \multirow{2}{*}{5} & Double & 7 & 5 & 6 & 5 & \multirow{2}{*}{7} & Double & 7 & 5 & 6 & 5\tabularnewline
\cline{3-7} \cline{9-13} 
 &  & Single & 1 & 1 & 1 & 1 &  & Single & 1 & 1 & 1 & 1\tabularnewline
\hline 
\multirow{2}{*}{RGM-56} & \multirow{2}{*}{5} & Double & 7 & 6 & 6 & 5 & \multirow{2}{*}{7} & Double & 7 & 6 & 6 & 5\tabularnewline
\cline{3-7} \cline{9-13} 
 &  & Single & 1 & 0 & 2 & 1 &  & Single & 1 & 0 & 2 & 1\tabularnewline
\hline 
\multirow{2}{*}{RGM-72} & \multirow{2}{*}{8} & Double & 9 & 7 & 9 & 7 & \multirow{2}{*}{8} & Double & 9 & 7 & 9 & 6\tabularnewline
\cline{3-7} \cline{9-13} 
 &  & Single & 1 & 1 & 1 & 1 &  & Single & 1 & 1 & 1 & 2\tabularnewline
\hline 
\end{tabular}
\end{table}

\section*{Comparison of Computed peaks Locations for RGMs with the Experiments
and Other Theoretical Calculations}

\begin{table}[H]
\caption{\label{tab:Comparison-of-bisanthene}Comparison of calculated peak
locations in the optical absorption spectra of RGM-28, with the experimental,
and other theoretical results, for bisanthene. The calculations were
performed using the PPP-MRSDCI approach, employing both the screened
and the standard parameters. All results are in eV units. }

\centering{}%
\begin{tabular}{|c|c|c|c|}
\hline 
\multirow{2}{*}{Experiments} & \multirow{2}{*}{Theory (others) } & \multicolumn{2}{c|}{This work}\tabularnewline
\cline{3-4} 
 &  & Screened & Standard\tabularnewline
\hline 
1.80\cite{Konishi_Bisanthenes},1.98\cite{Konishi_Bisanthenes},2.02\cite{clar1977corr}, & \href{http://www.dsf.unica.it/~gmalloci/pahs/bisanthene/bisanthene.html}{1.47}\cite{online_RQD28} & \multirow{2}{*}{2.00 $(^{1}B_{2u})$} & \multirow{2}{*}{2.21$(^{1}B_{2u})$}\tabularnewline
2.15\cite{arabei2000},2.43\cite{arabei2000}, & 1.78$^{a}$\cite{parac2003tddft}, 1.98$^{b}$\cite{parac2003tddft} &  & \tabularnewline
\hline 
- & 2.83$^{a}$\cite{parac2003tddft}, 2.96$^{b}$\cite{parac2003tddft}  & - & -\tabularnewline
\hline 
3.64\cite{Konishi_Bisanthenes},3.87\cite{arabei2000} & 3.76\cite{online_RQD28} & 3.87 $(^{1}B_{3u})$ & -\tabularnewline
\hline 
4.05\cite{Konishi_Bisanthenes} & - & 4.19$(^{1}B_{3u})$ & 4.14$(^{1}B_{2u}/^{1}B_{3u})$\tabularnewline
\hline 
4.80\cite{arabei2000} & 4.47\cite{online_RQD28} & 4.49$(^{1}B_{2u})$ & 4.68 $(^{1}B_{2u}/^{1}B_{3u})$\tabularnewline
\hline 
- & 5.39\cite{online_RQD28} & 5.22$(^{1}B_{2u}/^{1}B_{3u})$ & 5.09 $(^{1}B_{2u}/^{1}B_{3u})$\tabularnewline
\hline 
- & - & 5.63$(^{1}B_{3u})$ & 5.41$(^{1}B_{2u})$\tabularnewline
\hline 
- & - & 6.05$(^{1}B_{2u})$ & 6.06 $(^{1}B_{2u}/^{1}B_{3u})$\tabularnewline
\hline 
- & 6.35\cite{online_RQD28} & 6.41$(^{1}B_{2u})$ & -\tabularnewline
\hline 
- & 7.09\cite{online_RQD28} & 7.03$(^{1}B_{2u}/^{1}B_{3u})$ & 6.80$(^{1}B_{2u})$\tabularnewline
\hline 
\multicolumn{4}{|c|}{$^{a}$ TDDFT method, $^{b}$ TDPPP method}\tabularnewline
\hline 
\end{tabular}
\end{table}
\begin{table}[H]
\begin{centering}
\caption{\label{tab:Comparison-of-Terrylene}Comparison of calculated peak
locations in the optical absorption spectra of RGM-30, with the experimental,
and other theoretical results, for terrylene. The calculations were
performed using the PPP-MRSDCI approach, employing both the screened
and the standard parameters. All results are in eV units. }
\par\end{centering}
\centering{}%
\begin{tabular}{|c|c|c|c|}
\hline 
\multirow{2}{*}{Experiments} & \multirow{2}{*}{Theory (others) } & \multicolumn{2}{c|}{This work}\tabularnewline
\cline{3-4} 
 &  & Screened & Standard\tabularnewline
\hline 
2.14\cite{kummer1997absorption}, 2.21\cite{baumgarten1994spin,biktchantaev2002perylene},  & 2.02\cite{halasinski2003electronic}, 2.03\cite{viruela1992theoretical},  & \multirow{5}{*}{2.11\foreignlanguage{american}{ $(^{1}B_{2u})$}} & \multirow{5}{*}{2.43 \foreignlanguage{american}{$(^{1}B_{2u})$}}\tabularnewline
2.22\cite{koch1991polyarylenes}, 2.35\cite{ruiterkamp2002spectroscopy}, & 2.21$^{a}$/2.22$^{c}$\cite{malloci2011electronic}, 2.29\cite{minami2012theoretical}, &  & \tabularnewline
2.36\cite{clar1978correlations,halasinski2003electronic}, 2.39\cite{koch1991polyarylenes}, & 2.52\cite{karabunarliev1994structure}, 2.98\cite{halasinski2003electronic}, &  & \tabularnewline
2.57\cite{koch1991polyarylenes}, 2.76\cite{koch1991polyarylenes},  & 3.31\cite{halasinski2003electronic}, 3.40\cite{halasinski2003electronic},  &  & \tabularnewline
- & 3.84\cite{halasinski2003electronic},  &  & \tabularnewline
\hline 
- & - & 4.07 \foreignlanguage{american}{$(^{1}B_{3u})$} & -\tabularnewline
\hline 
4.33\cite{halasinski2003electronic}, 4.47\cite{koch1991polyarylenes} & - & 4.58 \foreignlanguage{american}{$(^{1}B_{3u})$} & 4.64\foreignlanguage{american}{ $(^{1}B_{2u})$}\tabularnewline
\hline 
4.71\cite{halasinski2003electronic,ruiterkamp2002spectroscopy} & 4.7\cite{malloci2011electronic} & - & \selectlanguage{american}%
4.75$(^{1}B_{3u})$\selectlanguage{english}%
\tabularnewline
\hline 
- & - & 5.12 \foreignlanguage{american}{$(^{1}B_{2u})$} & 5.03\foreignlanguage{american}{$(^{1}B_{3u})$}\tabularnewline
\hline 
5.20\cite{halasinski2003electronic}, 5.27\cite{ruiterkamp2002spectroscopy},  & - & \multirow{2}{*}{5.35 \foreignlanguage{american}{$(^{1}B_{3u})$}} & \multirow{2}{*}{5.53 \foreignlanguage{american}{$(^{1}B_{2u}/{}^{1}B_{3u})$}}\tabularnewline
5.41\cite{halasinski2003electronic}, 5.48\cite{koch1991polyarylenes} & - &  & \tabularnewline
\hline 
- & - & 5.95 \foreignlanguage{american}{$(^{1}B_{2u})$} & 5.87 \foreignlanguage{american}{$(^{1}B_{3u})$}\tabularnewline
\hline 
6.10\cite{koch1991polyarylenes}, 6.19\cite{halasinski2003electronic},  & 6\cite{malloci2011electronic} & 6.08 \foreignlanguage{american}{$(^{1}B_{3u})$} & 6.01 \foreignlanguage{american}{$(^{1}B_{2u})$}\tabularnewline
\hline 
6.44\cite{halasinski2003electronic},  & - & 6.47 $(^{1}B_{3u})$ & 6.24 \foreignlanguage{american}{$(^{1}B_{3u})$}\tabularnewline
\hline 
6.69\cite{halasinski2003electronic},  & 6.8\cite{malloci2011electronic} & 6.86 $(^{1}B_{3u})$ & 6.76 \foreignlanguage{american}{$(^{1}B_{3u})$}\tabularnewline
\hline 
\multicolumn{4}{|c|}{$^{a}$TDDFT method, $^{c}$DFT(Kohan-Sham) method}\tabularnewline
\hline 
\end{tabular}
\end{table}
\begin{table}[H]
\caption{\label{tab:Comparison-of-Tetrabenzocoronene}Comparison of calculated
peak locations in the optical absorption spectra of RGM-36, with theoretical
results of other authors for tetrabenzocoronene. No experimental results
are available for this molecule. The calculations were performed using
the PPP-MRSDCI approach, employing both the screened and the standard
parameters. All results are in eV units. }

\centering{}%
\begin{tabular}{|c|c|c|}
\hline 
\multirow{2}{*}{Theory (others)\cite{online_RQD36} } & \multicolumn{2}{c|}{This work}\tabularnewline
\cline{2-3} 
 & Screened & Standard\tabularnewline
\hline 
0.95 & \multirow{1}{*}{-} & \multirow{1}{*}{-}\tabularnewline
\hline 
- & 2.11 $(^{1}B_{2u})$ & 2.30$(^{1}B_{2u})$\tabularnewline
\hline 
3.16 & - & -\tabularnewline
\hline 
3.64 & 3.63$(^{1}B_{3u})$ & -\tabularnewline
\hline 
- & 3.77$(^{1}B_{2u})$ & 3.87$(^{1}B_{2u})$\tabularnewline
\hline 
- & 4.01$(^{1}B_{2u}/{}^{1}B_{3u})$ & -\tabularnewline
\hline 
- & 4.35$(^{1}B_{2u})$ & 4.41$(^{1}B_{2u}/{}^{1}B_{3u})$\tabularnewline
\hline 
4.83 & 4.94$(^{1}B_{3u})$ & 4.88$(^{1}B_{2u})$\tabularnewline
\hline 
- & - & 5.09$(^{1}B_{3u})$\tabularnewline
\hline 
- & 5.65$(^{1}B_{2u})$ & 5.60$(^{1}B_{3u})$\tabularnewline
\hline 
- & 5.99$(^{1}B_{2u}/{}^{1}B_{3u})$ & 5.86$(^{1}B_{2u}/{}^{1}B_{3u})$\tabularnewline
\hline 
6.22 & 6.50$(^{1}B_{2u})$ & 6.57$(^{1}B_{2u})$\tabularnewline
\hline 
\end{tabular}
\end{table}
\begin{table}[H]
\caption{\label{tab:Comparison-of-Quaterrylene}Comparison of calculated peak
locations in the optical absorption spectra of RGM-40, with the experimental,
and other theoretical results, for quaterrylene. The calculations
were performed using the PPP-MRSDCI approach, employing both the screened
and the standard parameters. All results are in eV units. }

\centering{}%
\begin{tabular}{|c|c|c|c|}
\hline 
\multirow{2}{*}{Experiment} & \multirow{2}{*}{Theory (others) } & \multicolumn{2}{c|}{This work}\tabularnewline
\cline{3-4} 
 &  & Screened & Standard\tabularnewline
\hline 
1.84\cite{former2002cyclodehydrogenation}, 1.87\cite{baumgarten1994spin,koch1991polyarylenes},  & 1.65\cite{viruela1992theoretical},1.67\cite{halasinski2003electronic}, & \multirow{6}{*}{2.02 $(^{1}B_{2u})$} & \multirow{6}{*}{2.30 $(^{1}B_{2u})$}\tabularnewline
1.91\cite{gudipati2006double}, 1.99\cite{former2002cyclodehydrogenation},  & 1.79$^{c}$/1.83$^{a}$\cite{malloci2011electronic}, &  & \tabularnewline
2.03\cite{clar1978correlations}, 2.04\cite{ruiterkamp2002spectroscopy,halasinski2003electronic}, & 1.87\cite{gudipati2006double}, 1.88\cite{minami2012theoretical}, &  & \tabularnewline
 & 2.18\cite{karabunarliev1994structure}, 2.97\cite{halasinski2003electronic}, &  & \tabularnewline
 & 3.13\cite{halasinski2003electronic}, 3.25\cite{halasinski2003electronic}, &  & \tabularnewline
 & 3.40\cite{halasinski2003electronic}, &  & \tabularnewline
\hline 
3.71\cite{koch1991polyarylenes}, 3.78 \cite{halasinski2003electronic},  & 3.60\cite{malloci2011electronic} & \multirow{2}{*}{4.06$(^{1}B_{2u}/{}^{1}B_{3u})$} & \multirow{2}{*}{-}\tabularnewline
3.85\cite{koch1991polyarylenes}, 3.86\cite{ruiterkamp2002spectroscopy}, &  &  & \tabularnewline
\hline 
- & 4.40\cite{malloci2011electronic} & 4.53$(^{1}B_{2u}/{}^{1}B_{3u})$ & 4.36 $(^{1}B_{2u})$\tabularnewline
\hline 
4.71\cite{halasinski2003electronic}, 4.83 \cite{halasinski2003electronic} & - & - & 4.85 $(^{1}B_{3u})$\tabularnewline
\hline 
5.27\cite{koch1991polyarylenes}, 5.39 \cite{halasinski2003electronic} & 5.30\cite{malloci2011electronic} & 5.16$(^{1}B_{2u}/{}^{1}B_{3u})$ & 5.48 $(^{1}B_{2u}/{}^{1}B_{3u})$\tabularnewline
\hline 
5.82 \cite{halasinski2003electronic} & - & 5.88$(^{1}B_{3u})$ & 5.85$(^{1}B_{3u})$\tabularnewline
\hline 
6.32 \cite{halasinski2003electronic} & 6.00\cite{malloci2011electronic} & 6.33$(^{1}B_{2u}/{}^{1}B_{3u})$ & 6.19 $(^{1}B_{2u}/{}^{1}B_{3u})$\tabularnewline
\hline 
6.50 \cite{halasinski2003electronic} & - & - & 6.45 $(^{1}B_{2u})$\tabularnewline
\hline 
6.63 \cite{halasinski2003electronic} & 6.60 \cite{malloci2011electronic} & 6.85$(^{1}B_{2u}/{}^{1}B_{3u})$ & 6.73 $(^{1}B_{3u})$\tabularnewline
\hline 
\multicolumn{4}{|c|}{$^{a}$TDDFT method, $^{c}$DFT(Kohan-Sham) method}\tabularnewline
\hline 
\end{tabular}
\end{table}
\begin{table}[H]
\caption{\label{tab:Comparison-of-Teranthene}Comparison of calculated peak
locations in the optical absorption spectra of RGM-42, with the experimental
results on t-butyl saturated teranthene. No other theoretical results
are available for this molecule. Our calculations were performed using
the PPP-MRSDCI approach, employing both the screened and the standard
parameters. All results are in eV units. }

\centering{}%
\begin{tabular}{|c|c|c|}
\hline 
\multirow{2}{*}{Experiment\cite{konishi2014anthenes} } & \multicolumn{2}{c|}{This work}\tabularnewline
\cline{2-3} 
 & Screened & Standard\tabularnewline
\hline 
1.17,1.21, & - & -\tabularnewline
\hline 
1.41, 1.57 & 1.86 $(^{1}B_{2u})$ & 2.04$(^{1}B_{2u})$\tabularnewline
\hline 
2.96 & - & -\tabularnewline
\hline 
3.19 & 3.56 $(^{1}B_{2u}/{}^{1}B_{3u})$ & -\tabularnewline
\hline 
3.87 & 3.96$(^{1}B_{2u})$ & 3.80$(^{1}B_{3u})$\tabularnewline
\hline 
- & - & 4.02 $(^{1}B_{2u})$\tabularnewline
\hline 
- & 4.15$(^{1}B_{2u})$ & 4.21 $(^{1}B_{3u})$\tabularnewline
\hline 
- & 4.53$(^{1}B_{3u})$ & 4.52$(^{1}B_{2u})$\tabularnewline
\hline 
\end{tabular}
\end{table}
\begin{table}[H]
\caption{\label{tab:Comparison-of-Pentarylene}Comparison of calculated peak
locations in the optical absorption spectra of RGM-50, with the experimental,
and other theoretical results, for pentarylene. The calculations were
performed using the PPP-MRSDCI approach, employing both the screened
and the standard parameters. All results are in eV units.}

\centering{}%
\begin{tabular}{|c|c|c|c|}
\hline 
\multirow{2}{*}{Experiments} & \multirow{2}{*}{Theory (others)} & \multicolumn{2}{c|}{This work}\tabularnewline
\cline{3-4} 
 &  & Screened & Standard\tabularnewline
\hline 
1.66\cite{baumgarten1994spin,koch1991polyarylenes}  & 1.40\cite{viruela1992theoretical}, 1.51$^{c}$/1.54$^{a}$\cite{malloci2011electronic}
,  & \multirow{3}{*}{1.72 $(^{1}B_{2u})$} & \multirow{3}{*}{1.98 $(^{1}B_{2u})$}\tabularnewline
 &  1.60\cite{minami2012theoretical}, &  & \tabularnewline
 & 1.97\cite{karabunarliev1994structure} &  & \tabularnewline
\hline 
3.28\cite{koch1991polyarylenes}, 3.45\cite{koch1991polyarylenes} & - & \multirow{1}{*}{3.39 $(^{1}B_{2u})$} & \multirow{1}{*}{-}\tabularnewline
\hline 
- & 4.0\cite{malloci2011electronic} & 3.91$(^{1}B_{2u}/{}^{1}B_{3u})$ & 3.84 $(^{1}B_{2u})$\tabularnewline
\hline 
- & - & 4.21$(^{1}B_{3u})$ & \tabularnewline
\hline 
4.62\cite{koch1991polyarylenes} & 4.5\cite{malloci2011electronic} & - & 4.71 $(^{1}B_{3u})$\tabularnewline
\hline 
4.80\cite{koch1991polyarylenes} & 5.2\cite{malloci2011electronic} & 4.97$(^{1}B_{2u}/{}^{1}B_{3u})$ & 5.12$(^{1}B_{2u}/{}^{1}B_{3u})$\tabularnewline
\hline 
5.29\cite{koch1991polyarylenes} & - & 5.41$(^{1}B_{3u})$ & 5.34 $(^{1}B_{3u})$\tabularnewline
\hline 
- & - & 5.73$(^{1}B_{3u})$ & 5.62 $(^{1}B_{2u}/{}^{1}B_{3u})$\tabularnewline
\hline 
- & 6.1\cite{malloci2011electronic} & 5.95$(^{1}B_{3u})$ & 5.99 $(^{1}B_{3u})$\tabularnewline
\hline 
- & - & 6.23$(^{1}B_{2u})$ & 6.41 $(^{1}B_{2u}/{}^{1}B_{3u})$\tabularnewline
\hline 
- & 7.4\cite{malloci2011electronic} & - & 6.96$(^{1}B_{3u})$\tabularnewline
\hline 
\multicolumn{4}{|c|}{ $^{a}$ TDDFT method, $^{c}$DFT(Kohan-Sham) method}\tabularnewline
\hline 
\end{tabular}
\end{table}
\begin{table}[H]
\caption{\label{peak_location_RGQD-54}The calculated peak locations in the
optical absorption spectra of RGM-54. No other theoretical and experimental
results are available for this molecule. Our calculations were performed
using the PPP-MRSDCI approach, employing both the screened and the
standard parameters. All results are in eV units. }

\centering{}%
\begin{tabular}{|c|c|}
\hline 
\multicolumn{2}{|c|}{This work}\tabularnewline
\hline 
Screened & Standard\tabularnewline
\hline 
1.63 $(^{1}B_{2u})$ & 2.09 $(^{1}B_{2u})$\tabularnewline
\hline 
2.56 $(^{1}B_{3u})$ & -\tabularnewline
\hline 
2.83 $(^{1}B_{2u})$ & -\tabularnewline
\hline 
3.09 $(^{1}B_{2u})$ & 3.20$(^{1}B_{2u}/{}^{1}B_{3u})$\tabularnewline
\hline 
3.71 $(^{1}B_{2u}/{}^{1}B_{3u})$ & 3.69 $(^{1}B_{2u})$\tabularnewline
\hline 
3.95 $(^{1}B_{3u})$ & 3.98 $(^{1}B_{2u}/{}^{1}B_{3u})$\tabularnewline
\hline 
4.15$(^{1}B_{2u})$ & 4.22$(^{1}B_{3u})$\tabularnewline
\hline 
4.31$(^{1}B_{2u})$ & 4.60 $(^{1}B_{3u})$\tabularnewline
\hline 
- & 4.97 $(^{1}B_{3u})$\tabularnewline
\hline 
5.14 $(^{1}B_{2u})$ & 5.14$(^{1}B_{2u})$\tabularnewline
\hline 
5.40 $(^{1}B_{3u})$ & 5.41 $(^{1}B_{3u})$\tabularnewline
\hline 
5.60 $(^{1}B_{3u})$ & -\tabularnewline
\hline 
5.82 $(^{1}B_{3u})$ & 5.97 $(^{1}B_{2u}/{}^{1}B_{3u})$\tabularnewline
\hline 
- & 6.22 $(^{1}B_{2u})$\tabularnewline
\hline 
- & 6.56 $(^{1}B_{2u}/{}^{1}B_{3u})$\tabularnewline
\hline 
\end{tabular}
\end{table}
\begin{table}[H]
\caption{\label{tab:Comparison-of-Quateranthene}Comparison of the calculated
peak locations in the optical absorption spectra of RGM-56, with the
experimental results of Konishi et al.\cite{konishi2014anthenes}
No other previous theoretical calculations of absorption spectrum
exist for this molecule. The calculations were performed using the
PPP-MRSDCI approach, employing both the screened and the standard
parameters. All results are in eV units.}

\centering{}%
\begin{tabular}{|c|c|c|}
\hline 
\multirow{2}{*}{Experiment\cite{konishi2014anthenes} } & \multicolumn{2}{c|}{This work}\tabularnewline
\cline{2-3} 
 & Screened & Standard\tabularnewline
\hline 
1.35, 2.01, 2.10, & \multirow{1}{*}{1.50 $(^{1}B_{2u})$} & \multirow{1}{*}{1.91$(^{1}B_{2u})$}\tabularnewline
\hline 
2.20, 2.27, 2.32 & \multirow{1}{*}{-} & \multirow{1}{*}{-}\tabularnewline
\hline 
- & 2.79 $(^{1}B_{2u}/{}^{1}B_{3u})$ & -\tabularnewline
\hline 
3.21 & - & 3.35$(^{1}B_{2u}/{}^{1}B_{3u})$\tabularnewline
\hline 
3.46 & 3.61$(^{1}B_{2u}/{}^{1}B_{3u})$ & -\tabularnewline
\hline 
- & 3.92$(^{1}B_{3u})$ & 3.87$(^{1}B_{2u})$\tabularnewline
\hline 
\end{tabular}
\end{table}

\section*{Detailed Information About the Excited States}

In the following tables, we present detailed information about the
excitation energies, dominant many-body wave-functions, and transition
dipole matrix elements of excited states with respect to the ground
state (1$^{1}$A$_{g}$). The coefficient of charge conjugate of a
given configuration is abbreviated as '$c.c.$' while the sign (+/-)
preceding '$c.c.$' indicates that the two coefficients have (same/opposite)
signs. Symbol $H$ denotes HOMO, while $L$ denotes LUMO. Similarly
$H-n$ and $L+m$ denote $n$-th orbital below HOMO, and $m$-th orbtial
above LUMO, respectively. The symbol $|H\rightarrow L\rangle$ denotes
a singly excited configuration obtained by promoting one electron
from HOMO to LUMO, with respect to the closed-shell Hartree-Fock reference
state. Similarly, one can deduce the meaning of other singly-excited
configurations such as $|H-1\rightarrow L+2\rangle$ \emph{etc}. The
symbol $|H\rightarrow L;H-1\rightarrow L+1\rangle$ denotes a doubly-excited
configuration obtained by exciting two electrons from the Hartree-Fock
reference state, one from HOMO to LUMO, the other one from HOMO-1
to LUMO+1. Nature of other doubly-excited configurations can also
be deduced, similarly.

\begin{table}[H]
\protect\caption{\label{wave_analysis_28atoms_scr_RQD}Excited states giving rise to
peaks in the singlet linear optical absorption spectrum of RGM-28,
computed employing the MRSDCI approach, along with the screened parameters
in the PPP model Hamiltonian. }

\selectlanguage{american}%
\centering{}%
\begin{tabular}{ccccc}
\hline 
\selectlanguage{english}%
Peak\selectlanguage{american}%
 & \selectlanguage{english}%
State \selectlanguage{american}%
 & \selectlanguage{english}%
E ($eV$)\selectlanguage{american}%
 & \selectlanguage{english}%
Transition \selectlanguage{american}%
 & \selectlanguage{english}%
Dominant Contributing \selectlanguage{american}%
\tabularnewline
\selectlanguage{english}%
\selectlanguage{american}%
 & \selectlanguage{english}%
\selectlanguage{american}%
 & \selectlanguage{english}%
\selectlanguage{american}%
 & \selectlanguage{english}%
Dipole (\AA)\selectlanguage{american}%
 & \selectlanguage{english}%
Configurations\selectlanguage{american}%
\tabularnewline
\hline 
$I_{y}$ & $1$$^{1}B_{2u}$ & $2.00$ & $1.5360$ & \selectlanguage{english}%
$\left|H\rightarrow L\right\rangle $ $(0.8548)$\selectlanguage{american}%
\tabularnewline
\selectlanguage{english}%
\selectlanguage{american}%
 & \selectlanguage{english}%
\selectlanguage{american}%
 & \selectlanguage{english}%
\selectlanguage{american}%
 & \selectlanguage{english}%
\selectlanguage{american}%
 & \selectlanguage{english}%
$\left|H\rightarrow L;H-2\rightarrow L\right\rangle $$-c.c.$ $(0.0772)$\selectlanguage{american}%
\tabularnewline
$II_{x}$ & $3$$^{1}B_{3u}$ & $3.87$ & $1.7616$ & \selectlanguage{english}%
$\left|H\rightarrow L+3\right\rangle $$+c.c.$ $(0.5731)$\selectlanguage{american}%
\tabularnewline
\selectlanguage{english}%
\selectlanguage{american}%
 & \selectlanguage{english}%
\selectlanguage{american}%
 & \selectlanguage{english}%
\selectlanguage{american}%
 & \selectlanguage{english}%
\selectlanguage{american}%
 & \selectlanguage{english}%
$\left|H\rightarrow L;H-4\rightarrow L\right\rangle $$+c.c.$ $(0.1326)$\selectlanguage{american}%
\tabularnewline
$III_{x}$ & $5$$^{1}B_{3u}$ & $4.19$ & $1.3240$ & \selectlanguage{english}%
$\left|H\rightarrow L;H\rightarrow L+1\right\rangle $$+c.c.$ $(0.4853)$\selectlanguage{american}%
\tabularnewline
\hline 
\end{tabular}\selectlanguage{english}%
\end{table}
\begin{table}[H]
\protect\caption{Excited states giving rise to peaks in the singlet linear optical
absorption spectrum of RGM-28, computed employing the MRSDCI approach,
along with the standard parameters in the PPP model Hamiltonian. \label{wave_analysis_28atoms_std_RQD}}

\selectlanguage{american}%
\centering{}%
\begin{tabular}{ccccc}
\hline 
\selectlanguage{english}%
Peak\selectlanguage{american}%
 & \selectlanguage{english}%
State \selectlanguage{american}%
 & \selectlanguage{english}%
E ($eV$)\selectlanguage{american}%
 & \selectlanguage{english}%
Transition \selectlanguage{american}%
 & \selectlanguage{english}%
Dominant Contributing \selectlanguage{american}%
\tabularnewline
\selectlanguage{english}%
\selectlanguage{american}%
 & \selectlanguage{english}%
\selectlanguage{american}%
 & \selectlanguage{english}%
\selectlanguage{american}%
 & \selectlanguage{english}%
Dipole (\AA)\selectlanguage{american}%
 & \selectlanguage{english}%
Configurations\selectlanguage{american}%
\tabularnewline
\hline 
$I_{y}$ & $1$$^{1}B_{2u}$ & $2.21$ & $1.4389$ & \selectlanguage{english}%
$\left|H\rightarrow L\right\rangle $ $(0.8536)$\selectlanguage{american}%
\tabularnewline
\selectlanguage{english}%
\selectlanguage{american}%
 & \selectlanguage{english}%
\selectlanguage{american}%
 & \selectlanguage{english}%
\selectlanguage{american}%
 & \selectlanguage{english}%
\selectlanguage{american}%
 & \selectlanguage{english}%
$\left|H\rightarrow L,H\rightarrow L+2\right\rangle $ $+c.c.$$(0.0849)$\selectlanguage{american}%
\tabularnewline
$II_{x\&y}$ & $4$$^{1}B_{2u}$ & $4.06$ & $0.3971$ & \selectlanguage{english}%
$\left|H\rightarrow L+5\right\rangle $ $-c.c.$$(0.5552)$\selectlanguage{american}%
\tabularnewline
\selectlanguage{english}%
\selectlanguage{american}%
 & \selectlanguage{english}%
\selectlanguage{american}%
 & \selectlanguage{english}%
\selectlanguage{american}%
 & \selectlanguage{english}%
\selectlanguage{american}%
 & \selectlanguage{english}%
$\left|H-1\rightarrow L+1\right\rangle $ $(0.2901)$\selectlanguage{american}%
\tabularnewline
\hline 
\end{tabular}\selectlanguage{english}%
\end{table}
\begin{table}[H]
\protect\caption{\label{wave_analysis_30atoms_scr_RQD}Excited states giving rise to
peaks in the singlet linear optical absorption spectrum of RGM-30,
computed employing the MRSDCI approach, along with the screened parameters
in the PPP model Hamiltonian. }

\selectlanguage{american}%
\centering{}%
\begin{tabular}{ccccc}
\hline 
\selectlanguage{english}%
Peak\selectlanguage{american}%
 & \selectlanguage{english}%
State \selectlanguage{american}%
 & \selectlanguage{english}%
E ($eV$)\selectlanguage{american}%
 & \selectlanguage{english}%
Transition \selectlanguage{american}%
 & \selectlanguage{english}%
Dominant Contributing \selectlanguage{american}%
\tabularnewline
\selectlanguage{english}%
\selectlanguage{american}%
 & \selectlanguage{english}%
\selectlanguage{american}%
 & \selectlanguage{english}%
\selectlanguage{american}%
 & \selectlanguage{english}%
Dipole (\AA)\selectlanguage{american}%
 & \selectlanguage{english}%
Configurations\selectlanguage{american}%
\tabularnewline
\hline 
$I_{y}$ & $1$$^{1}B_{2u}$ & $2.11$ & $2.3375$ & \selectlanguage{english}%
$\left|H\rightarrow L\right\rangle $ $(0.8651)$\selectlanguage{american}%
\tabularnewline
\selectlanguage{english}%
\selectlanguage{american}%
 & \selectlanguage{english}%
\selectlanguage{american}%
 & \selectlanguage{english}%
\selectlanguage{american}%
 & \selectlanguage{english}%
\selectlanguage{american}%
 & \selectlanguage{english}%
$\left|H\rightarrow L;H-1\rightarrow L\right\rangle $$-c.c.$ $(0.0599)$\selectlanguage{american}%
\tabularnewline
$II_{x}$ & $5$$^{1}B_{3u}$ & $4.07$ & $1.4797$ & \selectlanguage{english}%
$\left|H-3\rightarrow L\right\rangle $$-c.c.$$(0.5760)$\selectlanguage{american}%
\tabularnewline
\selectlanguage{english}%
\selectlanguage{american}%
 & \selectlanguage{english}%
\selectlanguage{american}%
 & \selectlanguage{english}%
\selectlanguage{american}%
 & \selectlanguage{english}%
\selectlanguage{american}%
 & \selectlanguage{english}%
$\left|H\rightarrow L;H-4\rightarrow L\right\rangle $$+c.c.$$(0.1246)$\selectlanguage{american}%
\tabularnewline
$III_{x}$ & $8$$^{1}B_{3u}$ & $4.58$ & $0.5733$ & \selectlanguage{english}%
$\left|H-7\rightarrow L\right\rangle $$-c.c.$ $(0.5012)$\selectlanguage{american}%
\tabularnewline
\hline 
\end{tabular}\selectlanguage{english}%
\end{table}
\begin{table}[H]
\protect\caption{\label{wave_analysis_30atoms_std_RQD}Excited states giving rise to
peaks in the singlet linear optical absorption spectrum of RGM-30,
computed employing the MRSDCI approach, along with the standard parameters
in the PPP model Hamiltonian.}

\selectlanguage{american}%
\centering{}%
\begin{tabular}{ccccc}
\hline 
\selectlanguage{english}%
Peak\selectlanguage{american}%
 & \selectlanguage{english}%
State \selectlanguage{american}%
 & \selectlanguage{english}%
E ($eV$)\selectlanguage{american}%
 & \selectlanguage{english}%
Transition \selectlanguage{american}%
 & \selectlanguage{english}%
Dominant Contributing \selectlanguage{american}%
\tabularnewline
\selectlanguage{english}%
\selectlanguage{american}%
 & \selectlanguage{english}%
\selectlanguage{american}%
 & \selectlanguage{english}%
\selectlanguage{american}%
 & \selectlanguage{english}%
Dipole (\AA)\selectlanguage{american}%
 & \selectlanguage{english}%
Configurations\selectlanguage{american}%
\tabularnewline
\hline 
$I_{y}$ & $1$$^{1}B_{2u}$ & $2.43$ & $2.1023$ & \selectlanguage{english}%
$\left|H\rightarrow L\right\rangle $ $(0.8590)$\selectlanguage{american}%
\tabularnewline
\selectlanguage{english}%
\selectlanguage{american}%
 & \selectlanguage{english}%
\selectlanguage{american}%
 & \selectlanguage{english}%
\selectlanguage{american}%
 & \selectlanguage{english}%
\selectlanguage{american}%
 & \selectlanguage{english}%
$\left|H-1\rightarrow L+1\right\rangle $ $(0.1237)$\selectlanguage{american}%
\tabularnewline
$II_{x\&y}$ & $3$$^{1}B_{2u}$ & $4.64$ & $0.3034$ & \selectlanguage{english}%
$\left|H\rightarrow L+6\right\rangle $$+c.c.$$(0.4912)$\selectlanguage{american}%
\tabularnewline
\selectlanguage{english}%
\selectlanguage{american}%
 & \selectlanguage{english}%
\selectlanguage{american}%
 & \selectlanguage{english}%
\selectlanguage{american}%
 & \selectlanguage{english}%
\selectlanguage{american}%
 & \selectlanguage{english}%
$\left|H-1\rightarrow L+1\right\rangle $ $(0.3289)$\selectlanguage{american}%
\tabularnewline
\selectlanguage{english}%
\selectlanguage{american}%
 & $6$$^{1}B_{3u}$ & $4.75$ & \selectlanguage{english}%
$1.0041$\selectlanguage{american}%
 & \selectlanguage{english}%
$\left|H\rightarrow L+3\right\rangle $$-c.c.$$(0.4511)$\selectlanguage{american}%
\tabularnewline
\hline 
\end{tabular}\selectlanguage{english}%
\end{table}
\begin{table}[H]
\protect\caption{\label{wave_analysis_36atoms_scr_RQD}Excited states giving rise to
peaks in the singlet linear optical absorption spectrum of RGM-36,
computed employing the MRSDCI approach, along with the screened parameters
in the PPP model. }

\selectlanguage{american}%
\centering{}%
\begin{tabular}{ccccc}
\hline 
\selectlanguage{english}%
Peak\selectlanguage{american}%
 & \selectlanguage{english}%
State \selectlanguage{american}%
 & \selectlanguage{english}%
E ($eV$)\selectlanguage{american}%
 & \selectlanguage{english}%
Transition \selectlanguage{american}%
 & \selectlanguage{english}%
Dominant Contributing \selectlanguage{american}%
\tabularnewline
\selectlanguage{english}%
\selectlanguage{american}%
 & \selectlanguage{english}%
\selectlanguage{american}%
 & \selectlanguage{english}%
\selectlanguage{american}%
 & \selectlanguage{english}%
Dipole (\AA)\selectlanguage{american}%
 & \selectlanguage{english}%
Configurations\selectlanguage{american}%
\tabularnewline
\hline 
$I_{y}$ & $1$$^{1}B_{2u}$ & $2.11$ & $1.5754$ & \selectlanguage{english}%
$\left|H\rightarrow L\right\rangle $ $(0.8489)$\selectlanguage{american}%
\tabularnewline
\selectlanguage{english}%
\selectlanguage{american}%
 & \selectlanguage{english}%
\selectlanguage{american}%
 & \selectlanguage{english}%
\selectlanguage{american}%
 & \selectlanguage{english}%
\selectlanguage{american}%
 & \selectlanguage{english}%
$\left|H-1\rightarrow L+1\right\rangle $ $(0.0983)$\selectlanguage{american}%
\tabularnewline
$II_{x}$ & $3$$^{1}B_{3u}$ & $3.63$ & $1.5116$ & \selectlanguage{english}%
$\left|H\rightarrow L;H\rightarrow L+1\right\rangle $$-c.c.$ $(0.5303)$\selectlanguage{american}%
\tabularnewline
\selectlanguage{english}%
\selectlanguage{american}%
 & \selectlanguage{english}%
\selectlanguage{american}%
 & \selectlanguage{english}%
\selectlanguage{american}%
 & \selectlanguage{english}%
\selectlanguage{american}%
 & \selectlanguage{english}%
$\left|H\rightarrow L+6\right\rangle $$+c.c.$$(0.1512)$\selectlanguage{american}%
\tabularnewline
$III_{y}$ & $4$$^{1}B_{2u}$ & $3.77$ & $1.4492$ & \selectlanguage{english}%
$\left|H-1\rightarrow L+1\right\rangle $ $(0.6509)$\selectlanguage{american}%
\tabularnewline
\hline 
\end{tabular}\selectlanguage{english}%
\end{table}
\begin{table}[H]
\protect\caption{\label{wave_analysis_36atoms_std_RQD}Excited states giving rise to
peaks in the singlet linear optical absorption spectrum of RGM-36,
computed employing the MRSDCI approach, along with the standard parameters
in the PPP model. }

\selectlanguage{american}%
\centering{}%
\begin{tabular}{ccccc}
\hline 
\selectlanguage{english}%
Peak\selectlanguage{american}%
 & \selectlanguage{english}%
State \selectlanguage{american}%
 & \selectlanguage{english}%
E ($eV$)\selectlanguage{american}%
 & \selectlanguage{english}%
Transition \selectlanguage{american}%
 & \selectlanguage{english}%
Dominant Contributing \selectlanguage{american}%
\tabularnewline
\selectlanguage{english}%
\selectlanguage{american}%
 & \selectlanguage{english}%
\selectlanguage{american}%
 & \selectlanguage{english}%
\selectlanguage{american}%
 & \selectlanguage{english}%
Dipole (\AA)\selectlanguage{american}%
 & \selectlanguage{english}%
Configurations\selectlanguage{american}%
\tabularnewline
\hline 
$I_{y}$ & $1$$^{1}B_{2u}$ & $2.30$ & $1.406716$ & \selectlanguage{english}%
$\left|H\rightarrow L\right\rangle $ $(0.8352)$\selectlanguage{american}%
\tabularnewline
\selectlanguage{english}%
\selectlanguage{american}%
 & \selectlanguage{english}%
\selectlanguage{american}%
 & \selectlanguage{english}%
\selectlanguage{american}%
 & \selectlanguage{english}%
\selectlanguage{american}%
 & \selectlanguage{english}%
$\left|H-1\rightarrow L+1\right\rangle $ $(0.1018)$\selectlanguage{american}%
\tabularnewline
$II_{y}$ & $4$$^{1}B_{2u}$ & $3.87$ & $0.913194$ & \selectlanguage{english}%
$\left|H-1\rightarrow L+1\right\rangle $ $(0.4670)$\selectlanguage{american}%
\tabularnewline
\selectlanguage{english}%
\selectlanguage{american}%
 & \selectlanguage{english}%
\selectlanguage{american}%
 & \selectlanguage{english}%
\selectlanguage{american}%
 & \selectlanguage{english}%
\selectlanguage{american}%
 & \selectlanguage{english}%
$\left|H-4\rightarrow L\right\rangle $$-c.c.$$(0.4507)$\selectlanguage{american}%
\tabularnewline
$III_{x\&y}$ & $5$$^{1}B_{2u}$ & $4.35$ & $0.727174$ & \selectlanguage{english}%
$\left|H-1\rightarrow L+1\right\rangle $ $(0.4166)$\selectlanguage{american}%
\tabularnewline
\hline 
\end{tabular}\selectlanguage{english}%
\end{table}
\begin{table}[H]
\protect\caption{\label{wave_analysis_40atoms_scr_RQD}Excited states giving rise to
peaks in the singlet linear optical absorption spectrum of RGM-40,
computed employing the MRSDCI approach, along with the screened parameters
in the PPP model. }

\selectlanguage{american}%
\centering{}%
\begin{tabular}{ccccc}
\hline 
\selectlanguage{english}%
Peak\selectlanguage{american}%
 & \selectlanguage{english}%
State \selectlanguage{american}%
 & \selectlanguage{english}%
E ($eV$)\selectlanguage{american}%
 & \selectlanguage{english}%
Transition \selectlanguage{american}%
 & \selectlanguage{english}%
Dominant Contributing \selectlanguage{american}%
\tabularnewline
\selectlanguage{english}%
\selectlanguage{american}%
 & \selectlanguage{english}%
\selectlanguage{american}%
 & \selectlanguage{english}%
\selectlanguage{american}%
 & \selectlanguage{english}%
Dipole (\AA)\selectlanguage{american}%
 & \selectlanguage{english}%
Configurations\selectlanguage{american}%
\tabularnewline
\hline 
$I_{y}$ & $1$$^{1}B_{2u}$ & $2.02$ & $3.1106$ & \selectlanguage{english}%
$\left|H\rightarrow L\right\rangle $ $(0.8557)$\selectlanguage{american}%
\tabularnewline
\selectlanguage{english}%
\selectlanguage{american}%
 & \selectlanguage{english}%
\selectlanguage{american}%
 & \selectlanguage{english}%
\selectlanguage{american}%
 & \selectlanguage{english}%
\selectlanguage{american}%
 & \selectlanguage{english}%
$\left|H\rightarrow L;H-1\rightarrow L\right\rangle $$-c.c.$ $(0.0659)$\selectlanguage{american}%
\tabularnewline
$II_{x\&y}$ & $4$$^{1}B_{2u}$ & $3.97$ & $0.7433$ & \selectlanguage{english}%
$\left|H-1\rightarrow L+1\right\rangle $$(0.5418)$\selectlanguage{american}%
\tabularnewline
\selectlanguage{english}%
\selectlanguage{american}%
 & \selectlanguage{english}%
\selectlanguage{american}%
 & \selectlanguage{english}%
\selectlanguage{american}%
 & \selectlanguage{english}%
\selectlanguage{american}%
 & \selectlanguage{english}%
$\left|H\rightarrow L+3\right\rangle $$+c.c.$$(0.4542)$\selectlanguage{american}%
\tabularnewline
\hline 
\end{tabular}\selectlanguage{english}%
\end{table}
\begin{table}[H]
\protect\caption{\label{wave_analysis_40atoms_std_RQD}Excited states giving rise to
peaks in the singlet linear optical absorption spectrum of RGM-40,
computed employing the MRSDCI approach, along with the standard parameters
in the PPP model. }

\selectlanguage{american}%
\centering{}%
\begin{tabular}{ccccc}
\hline 
\selectlanguage{english}%
Peak\selectlanguage{american}%
 & \selectlanguage{english}%
State \selectlanguage{american}%
 & \selectlanguage{english}%
E ($eV$)\selectlanguage{american}%
 & \selectlanguage{english}%
Transition \selectlanguage{american}%
 & \selectlanguage{english}%
Dominant Contributing \selectlanguage{american}%
\tabularnewline
\selectlanguage{english}%
\selectlanguage{american}%
 & \selectlanguage{english}%
\selectlanguage{american}%
 & \selectlanguage{english}%
\selectlanguage{american}%
 & \selectlanguage{english}%
Dipole (\AA)\selectlanguage{american}%
 & \selectlanguage{english}%
Configurations\selectlanguage{american}%
\tabularnewline
\hline 
$I_{y}$ & $1$$^{1}B_{2u}$ & $2.30$ & $2.7613$ & \selectlanguage{english}%
$\left|H\rightarrow L\right\rangle $ $(0.8402)$\selectlanguage{american}%
\tabularnewline
\selectlanguage{english}%
\selectlanguage{american}%
 & \selectlanguage{english}%
\selectlanguage{american}%
 & \selectlanguage{english}%
\selectlanguage{american}%
 & \selectlanguage{english}%
\selectlanguage{american}%
 & \selectlanguage{english}%
$\left|H-1\rightarrow L+1\right\rangle $ $(0.1470)$\selectlanguage{american}%
\tabularnewline
$II_{y}$ & $3$$^{1}B_{2u}$ & $4.36$ & $0.5470$ & \selectlanguage{english}%
$\left|H\rightarrow L+3\right\rangle $$+c.c.$$(0.5117)$\selectlanguage{american}%
\tabularnewline
\selectlanguage{english}%
\selectlanguage{american}%
 & \selectlanguage{english}%
\selectlanguage{american}%
 & \selectlanguage{english}%
\selectlanguage{american}%
 & \selectlanguage{english}%
\selectlanguage{american}%
 & \selectlanguage{english}%
$\left|H-1\rightarrow L+1\right\rangle $ $(0.3179)$\selectlanguage{american}%
\tabularnewline
$III_{x}$ & $6$$^{1}B_{3u}$ & $4.85$ & $0.8881$ & \selectlanguage{english}%
$\left|H-4\rightarrow L\right\rangle $$+c.c.$ $(0.3945)$\selectlanguage{american}%
\tabularnewline
\hline 
\end{tabular}\selectlanguage{english}%
\end{table}
\begin{table}[H]
\protect\caption{\label{wave_analysis_42atoms_scr_RQD}Excited states giving rise to
peaks in the singlet linear optical absorption spectrum of RGM-42,
computed employing the MRSDCI approach, along with the screened parameters
in the PPP model. }

\selectlanguage{american}%
\centering{}%
\begin{tabular}{ccccc}
\hline 
\selectlanguage{english}%
Peak\selectlanguage{american}%
 & \selectlanguage{english}%
State \selectlanguage{american}%
 & \selectlanguage{english}%
E ($eV$)\selectlanguage{american}%
 & \selectlanguage{english}%
Transition \selectlanguage{american}%
 & \selectlanguage{english}%
Dominant Contributing \selectlanguage{american}%
\tabularnewline
\selectlanguage{english}%
\selectlanguage{american}%
 & \selectlanguage{english}%
\selectlanguage{american}%
 & \selectlanguage{english}%
\selectlanguage{american}%
 & \selectlanguage{english}%
Dipole (\AA)\selectlanguage{american}%
 & \selectlanguage{english}%
Configurations\selectlanguage{american}%
\tabularnewline
\hline 
$I_{y}$ & $1$$^{1}B_{2u}$ & $1.86$ & $2.5210$ & \selectlanguage{english}%
$\left|H\rightarrow L\right\rangle $ $(0.8412)$\selectlanguage{american}%
\tabularnewline
\selectlanguage{english}%
\selectlanguage{american}%
 & \selectlanguage{english}%
\selectlanguage{american}%
 & \selectlanguage{english}%
\selectlanguage{american}%
 & \selectlanguage{english}%
\selectlanguage{american}%
 & \selectlanguage{english}%
$\left|H\rightarrow L;H-2\rightarrow L\right\rangle $$-c.c.$ $(0.1094)$\selectlanguage{american}%
\tabularnewline
$II_{x\&y}$ & $4$$^{1}B_{3u}$ & $3.50$ & $1.2044$ & \selectlanguage{english}%
$\left|H\rightarrow L;H\rightarrow L+1\right\rangle $$+c.c.$ $(0.4984)$\selectlanguage{american}%
\tabularnewline
\selectlanguage{english}%
\selectlanguage{american}%
 & \selectlanguage{english}%
\selectlanguage{american}%
 & \selectlanguage{english}%
\selectlanguage{american}%
 & \selectlanguage{english}%
\selectlanguage{american}%
 & \selectlanguage{english}%
$\left|H-6\rightarrow L\right\rangle $$+c.c.$$(0.2112)$\selectlanguage{american}%
\tabularnewline
\hline 
\end{tabular}\selectlanguage{english}%
\end{table}
\begin{table}[H]
\protect\caption{\label{wave_analysis_42atoms_std_RQD}Excited states giving rise to
peaks in the singlet linear optical absorption spectrum of RGM-42,
computed employing the MRSDCI approach, along with the standard parameters
in the PPP model. }

\selectlanguage{american}%
\centering{}%
\begin{tabular}{ccccc}
\hline 
\selectlanguage{english}%
Peak\selectlanguage{american}%
 & \selectlanguage{english}%
State \selectlanguage{american}%
 & \selectlanguage{english}%
E ($eV$)\selectlanguage{american}%
 & \selectlanguage{english}%
Transition \selectlanguage{american}%
 & \selectlanguage{english}%
Dominant Contributing \selectlanguage{american}%
\tabularnewline
\selectlanguage{english}%
\selectlanguage{american}%
 & \selectlanguage{english}%
\selectlanguage{american}%
 & \selectlanguage{english}%
\selectlanguage{american}%
 & \selectlanguage{english}%
Dipole (\AA)\selectlanguage{american}%
 & \selectlanguage{english}%
Configurations\selectlanguage{american}%
\tabularnewline
\hline 
$I_{y}$ & $1$$^{1}B_{2u}$ & $2.04$ & $2.2700$ & \selectlanguage{english}%
$\left|H\rightarrow L\right\rangle $ $(0.8368)$\selectlanguage{american}%
\tabularnewline
\selectlanguage{english}%
\selectlanguage{american}%
 & \selectlanguage{english}%
\selectlanguage{american}%
 & \selectlanguage{english}%
\selectlanguage{american}%
 & \selectlanguage{english}%
\selectlanguage{american}%
 & \selectlanguage{english}%
$\left|H\rightarrow L,H\rightarrow L+1\right\rangle $ $-c.c.$$(0.1091)$\selectlanguage{american}%
\tabularnewline
$II_{x}$ & $4$$^{1}B_{3u}$ & $3.80$ & $0.4640$ & \selectlanguage{english}%
$\left|H\rightarrow L,H-2\rightarrow L\right\rangle $ $+c.c.$$(0.3563)$\selectlanguage{american}%
\tabularnewline
\selectlanguage{english}%
\selectlanguage{american}%
 & \selectlanguage{english}%
\selectlanguage{american}%
 & \selectlanguage{english}%
\selectlanguage{american}%
 & \selectlanguage{english}%
\selectlanguage{american}%
 & \selectlanguage{english}%
$\left|H\rightarrow L+6\right\rangle $$+c.c.$ $(0.3059)$\selectlanguage{american}%
\tabularnewline
$III_{y}$ & $5$$^{1}B_{2u}$ & $4.02$ & $0.8072$ & \selectlanguage{english}%
$\left|H\rightarrow L+8\right\rangle $$-c.c.$ $(0.3624)$\selectlanguage{american}%
\tabularnewline
\hline 
\end{tabular}\selectlanguage{english}%
\end{table}
\begin{table}[H]
\protect\caption{\label{wave_analysis_50atoms_scr_RQD}Excited states giving rise to
peaks in the singlet linear optical absorption spectrum of RGM-50,
computed employing the MRSDCI approach, along with the screened parameters
in the PPP model. }

\selectlanguage{american}%
\centering{}%
\begin{tabular}{ccccc}
\hline 
\selectlanguage{english}%
Peak\selectlanguage{american}%
 & \selectlanguage{english}%
State \selectlanguage{american}%
 & \selectlanguage{english}%
E ($eV$)\selectlanguage{american}%
 & \selectlanguage{english}%
Transition \selectlanguage{american}%
 & \selectlanguage{english}%
Dominant Contributing \selectlanguage{american}%
\tabularnewline
\selectlanguage{english}%
\selectlanguage{american}%
 & \selectlanguage{english}%
\selectlanguage{american}%
 & \selectlanguage{english}%
\selectlanguage{american}%
 & \selectlanguage{english}%
Dipole (\AA)\selectlanguage{american}%
 & \selectlanguage{english}%
Configurations\selectlanguage{american}%
\tabularnewline
\hline 
$I_{y}$ & $1$$^{1}B_{2u}$ & $1.72$ & $3.8432$ & \selectlanguage{english}%
$\left|H\rightarrow L\right\rangle $ $(0.8590)$\selectlanguage{american}%
\tabularnewline
\selectlanguage{english}%
\selectlanguage{american}%
 & \selectlanguage{english}%
\selectlanguage{american}%
 & \selectlanguage{english}%
\selectlanguage{american}%
 & \selectlanguage{english}%
\selectlanguage{american}%
 & \selectlanguage{english}%
$\left|H\rightarrow L;H\rightarrow L+1\right\rangle $$+c.c.$ $(0.0678)$\selectlanguage{american}%
\tabularnewline
$II_{y}$ & $4$$^{1}B_{2u}$ & $3.39$ & $0.9517$ & \selectlanguage{english}%
$\left|H-1\rightarrow L+1\right\rangle $$(0.5734)$\selectlanguage{american}%
\tabularnewline
\selectlanguage{english}%
\selectlanguage{american}%
 & \selectlanguage{english}%
\selectlanguage{american}%
 & \selectlanguage{english}%
\selectlanguage{american}%
 & \selectlanguage{english}%
\selectlanguage{american}%
 & \selectlanguage{english}%
$\left|H-2\rightarrow L\right\rangle $$-c.c.$$(0.4401)$\selectlanguage{american}%
\tabularnewline
$III_{x\&y}$ & $8$$^{1}B_{2u}$ & $3.79$ & $0.5483$ & \selectlanguage{english}%
$\left|H\text{\ensuremath{\rightarrow}}L;H\rightarrow L+1\right\rangle $
$+c.c.$ $(0.4451)$\selectlanguage{american}%
\tabularnewline
\hline 
\end{tabular}\selectlanguage{english}%
\end{table}
\begin{table}[H]
\protect\caption{\label{wave_analysis_50atoms_std_RQD}Excited states giving rise to
peaks in the singlet linear optical absorption spectrum of RGM-50,
computed employing the MRSDCI approach, along with the standard parameters
in the PPP model.}

\selectlanguage{american}%
\centering{}%
\begin{tabular}{ccccc}
\hline 
\selectlanguage{english}%
Peak\selectlanguage{american}%
 & \selectlanguage{english}%
State \selectlanguage{american}%
 & \selectlanguage{english}%
E ($eV$)\selectlanguage{american}%
 & \selectlanguage{english}%
Transition \selectlanguage{american}%
 & \selectlanguage{english}%
Dominant Contributing \selectlanguage{american}%
\tabularnewline
\selectlanguage{english}%
\selectlanguage{american}%
 & \selectlanguage{english}%
\selectlanguage{american}%
 & \selectlanguage{english}%
\selectlanguage{american}%
 & \selectlanguage{english}%
Dipole (\AA)\selectlanguage{american}%
 & \selectlanguage{english}%
Configurations\selectlanguage{american}%
\tabularnewline
\hline 
$I_{y}$ & $1$$^{1}B_{2u}$ & $1.98$ & $3.4780$ & \selectlanguage{english}%
$\left|H\rightarrow L\right\rangle $ $(0.8368)$\selectlanguage{american}%
\tabularnewline
\selectlanguage{english}%
\selectlanguage{american}%
 & \selectlanguage{english}%
\selectlanguage{american}%
 & \selectlanguage{english}%
\selectlanguage{american}%
 & \selectlanguage{english}%
\selectlanguage{american}%
 & \selectlanguage{english}%
$\left|H-1\rightarrow L+1\right\rangle $ $(0.1685)$\selectlanguage{american}%
\tabularnewline
$II_{y}$ & $4$$^{1}B_{2u}$ & $3.84$ & $0.6241$ & \selectlanguage{english}%
$\left|H-1\rightarrow L+1\right\rangle $ $(0.6657)$\selectlanguage{american}%
\tabularnewline
\selectlanguage{english}%
\selectlanguage{american}%
 & \selectlanguage{english}%
\selectlanguage{american}%
 & \selectlanguage{english}%
\selectlanguage{american}%
 & \selectlanguage{english}%
\selectlanguage{american}%
 & \selectlanguage{english}%
$\left|H-2\rightarrow L\right\rangle $$+c.c.$ $(0.2916)$\selectlanguage{american}%
\tabularnewline
$III_{x}$ & $8$$^{1}B_{3u}$ & $4.71$ & $0.8499$ & \selectlanguage{english}%
$\left|H-4\rightarrow L\right\rangle $$+c.c.$ $(0.3818)$\selectlanguage{american}%
\tabularnewline
\hline 
\end{tabular}\selectlanguage{english}%
\end{table}
\begin{table}[H]
\protect\caption{\label{wave_analysis_54atoms_scr_RQD}Excited states giving rise to
peaks in the singlet linear optical absorption spectrum of RGM-54,
computed employing the MRSDCI approach, along with the screened parameters
in the PPP model. }

\selectlanguage{american}%
\centering{}%
\begin{tabular}{ccccc}
\hline 
\selectlanguage{english}%
Peak\selectlanguage{american}%
 & \selectlanguage{english}%
State \selectlanguage{american}%
 & \selectlanguage{english}%
E ($eV$)\selectlanguage{american}%
 & \selectlanguage{english}%
Transition \selectlanguage{american}%
 & \selectlanguage{english}%
Dominant Contributing \selectlanguage{american}%
\tabularnewline
\selectlanguage{english}%
\selectlanguage{american}%
 & \selectlanguage{english}%
\selectlanguage{american}%
 & \selectlanguage{english}%
\selectlanguage{american}%
 & \selectlanguage{english}%
Dipole (\AA)\selectlanguage{american}%
 & \selectlanguage{english}%
Configurations\selectlanguage{american}%
\tabularnewline
\hline 
$I_{y}$ & $1$$^{1}B_{2u}$ & $1.63$ & $1.4845$ & \selectlanguage{english}%
$\left|H\rightarrow L\right\rangle $ $(0.8330)$\selectlanguage{american}%
\tabularnewline
\selectlanguage{english}%
\selectlanguage{american}%
 & \selectlanguage{english}%
\selectlanguage{american}%
 & \selectlanguage{english}%
\selectlanguage{american}%
 & \selectlanguage{english}%
\selectlanguage{american}%
 & \selectlanguage{english}%
$\left|H-1\rightarrow L+1\right\rangle $ $(0.1286)$\selectlanguage{american}%
\tabularnewline
$II_{x}$ & $2$$^{1}B_{3u}$ & $2.56$ & $2.2031$ & \selectlanguage{english}%
$\left|H\rightarrow L;H\rightarrow L+1\right\rangle $$+c.c.$ $(0.4984)$\selectlanguage{american}%
\tabularnewline
\selectlanguage{english}%
\selectlanguage{american}%
 & \selectlanguage{english}%
\selectlanguage{american}%
 & \selectlanguage{english}%
\selectlanguage{american}%
 & \selectlanguage{english}%
\selectlanguage{american}%
 & \selectlanguage{english}%
$\left|H-4\rightarrow L\right\rangle $$-c.c.$$(0.1687)$\selectlanguage{american}%
\tabularnewline
$III_{y}$ & $4$$^{1}B_{2u}$ & $2.83$ & $2.0444$ & \selectlanguage{english}%
$\left|H-1\rightarrow L+1\right\rangle $ $(0.6233)$\selectlanguage{american}%
\tabularnewline
\hline 
\end{tabular}\selectlanguage{english}%
\end{table}
\begin{table}[H]
\protect\caption{\label{wave_analysis_54atoms_std_RQD}Excited states giving rise to
peaks in the singlet linear optical absorption spectrum of RGM-54,
computed employing the MRSDCI approach, along with the standard parameters
in the PPP model. }

\selectlanguage{american}%
\centering{}%
\begin{tabular}{ccccc}
\hline 
\selectlanguage{english}%
Peak\selectlanguage{american}%
 & \selectlanguage{english}%
State \selectlanguage{american}%
 & \selectlanguage{english}%
E ($eV$)\selectlanguage{american}%
 & \selectlanguage{english}%
Transition \selectlanguage{american}%
 & \selectlanguage{english}%
Dominant Contributing \selectlanguage{american}%
\tabularnewline
\selectlanguage{english}%
\selectlanguage{american}%
 & \selectlanguage{english}%
\selectlanguage{american}%
 & \selectlanguage{english}%
\selectlanguage{american}%
 & \selectlanguage{english}%
Dipole (\AA)\selectlanguage{american}%
 & \selectlanguage{english}%
Configurations\selectlanguage{american}%
\tabularnewline
\hline 
$I_{y}$ & $1$$^{1}B_{2u}$ & $2.09$ & $1.4091$ & \selectlanguage{english}%
$\left|H\rightarrow L\right\rangle $ $(0.8149)$\selectlanguage{american}%
\tabularnewline
\selectlanguage{english}%
\selectlanguage{american}%
 & \selectlanguage{english}%
\selectlanguage{american}%
 & \selectlanguage{english}%
\selectlanguage{american}%
 & \selectlanguage{english}%
\selectlanguage{american}%
 & \selectlanguage{english}%
$\left|H-1\rightarrow L+1\right\rangle $ $(0.1651)$\selectlanguage{american}%
\tabularnewline
$II_{x\&y}$ & $2$$^{1}B_{3u}$ & $3.15$ & $1.6323$ & \selectlanguage{english}%
$\left|H\rightarrow L;H\rightarrow L+1\right\rangle $$-c.c.$ $(0.4840)$\selectlanguage{american}%
\tabularnewline
\selectlanguage{english}%
\selectlanguage{american}%
 & \selectlanguage{english}%
\selectlanguage{american}%
 & \selectlanguage{english}%
\selectlanguage{american}%
 & \selectlanguage{english}%
\selectlanguage{american}%
 & \selectlanguage{english}%
$\left|H-4\rightarrow L\right\rangle $$-c.c.$$(0.2369)$\selectlanguage{american}%
\tabularnewline
\hline 
\end{tabular}\selectlanguage{english}%
\end{table}
\begin{table}[H]
\protect\caption{\label{wave_analysis_56atoms_scr_RQD}Excited states giving rise to
peaks in the singlet linear optical absorption spectrum of RGM-56,
computed employing the MRSDCI approach, along with the screened parameters
in the PPP model. }

\selectlanguage{american}%
\centering{}%
\begin{tabular}{ccccc}
\hline 
\selectlanguage{english}%
Peak\selectlanguage{american}%
 & \selectlanguage{english}%
State \selectlanguage{american}%
 & \selectlanguage{english}%
E ($eV$)\selectlanguage{american}%
 & \selectlanguage{english}%
Transition \selectlanguage{american}%
 & \selectlanguage{english}%
Dominant Contributing \selectlanguage{american}%
\tabularnewline
\selectlanguage{english}%
\selectlanguage{american}%
 & \selectlanguage{english}%
\selectlanguage{american}%
 & \selectlanguage{english}%
\selectlanguage{american}%
 & \selectlanguage{english}%
Dipole (\AA)\selectlanguage{american}%
 & \selectlanguage{english}%
Configurations\selectlanguage{american}%
\tabularnewline
\hline 
$I_{y}$ & $1$$^{1}B_{2u}$ & $1.50$ & $2.7694$ & \selectlanguage{english}%
$\left|H\rightarrow L\right\rangle $ $(0.8452)$\selectlanguage{american}%
\tabularnewline
\selectlanguage{english}%
\selectlanguage{american}%
 & \selectlanguage{english}%
\selectlanguage{american}%
 & \selectlanguage{english}%
\selectlanguage{american}%
 & \selectlanguage{english}%
\selectlanguage{american}%
 & \selectlanguage{english}%
$\left|H\rightarrow L;H\rightarrow L+1\right\rangle $$-c.c.$ $(0.1424)$\selectlanguage{american}%
\tabularnewline
$II_{x\&y}$ & $3$$^{1}B_{2u}$ & $2.76$ & $2.2125$ & \selectlanguage{english}%
$\left|H\rightarrow L;H\rightarrow L+1\right\rangle $$-c.c.$ $(0.5010)$\selectlanguage{american}%
\tabularnewline
\selectlanguage{english}%
\selectlanguage{american}%
 & \selectlanguage{english}%
\selectlanguage{american}%
 & \selectlanguage{english}%
\selectlanguage{american}%
 & \selectlanguage{english}%
\selectlanguage{american}%
 & \selectlanguage{english}%
$\left|H-1\rightarrow L\right\rangle $$+c.c.$$(0.2824)$\selectlanguage{american}%
\tabularnewline
\hline 
\end{tabular}\selectlanguage{english}%
\end{table}
\begin{table}[H]
\protect\caption{\label{wave_analysis_56atoms_std_RQD}Excited states giving rise to
peaks in the singlet linear optical absorption spectrum of RGM-56,
computed employing the MRSDCI approach, along with the standard parameters
in the PPP model.}

\selectlanguage{american}%
\centering{}%
\begin{tabular}{ccccc}
\hline 
\selectlanguage{english}%
Peak\selectlanguage{american}%
 & \selectlanguage{english}%
State \selectlanguage{american}%
 & \selectlanguage{english}%
E ($eV$)\selectlanguage{american}%
 & \selectlanguage{english}%
Transition \selectlanguage{american}%
 & \selectlanguage{english}%
Dominant Contributing \selectlanguage{american}%
\tabularnewline
\selectlanguage{english}%
\selectlanguage{american}%
 & \selectlanguage{english}%
\selectlanguage{american}%
 & \selectlanguage{english}%
\selectlanguage{american}%
 & \selectlanguage{english}%
Dipole (\AA)\selectlanguage{american}%
 & \selectlanguage{english}%
Configurations\selectlanguage{american}%
\tabularnewline
\hline 
$I_{y}$ & $1$$^{1}B_{2u}$ & $1.91$ & $2.8248$ & \selectlanguage{english}%
$\left|H\rightarrow L\right\rangle $ $(0.8337)$\selectlanguage{american}%
\tabularnewline
\selectlanguage{english}%
\selectlanguage{american}%
 & \selectlanguage{english}%
\selectlanguage{american}%
 & \selectlanguage{english}%
\selectlanguage{american}%
 & \selectlanguage{english}%
\selectlanguage{american}%
 & \selectlanguage{english}%
$\left|H\rightarrow L;H\rightarrow L+1\right\rangle $$-c.c.$ $(0.1496)$\selectlanguage{american}%
\tabularnewline
$II_{x\&y}$ & $3$$^{1}B_{2u}$ & $3.32$ & $1.1646$ & \selectlanguage{english}%
$\left|H\rightarrow L;H\rightarrow L+1\right\rangle $$-c.c.$ $(0.3783)$\selectlanguage{american}%
\tabularnewline
\selectlanguage{english}%
\selectlanguage{american}%
 & \selectlanguage{english}%
\selectlanguage{american}%
 & \selectlanguage{english}%
\selectlanguage{american}%
 & \selectlanguage{english}%
\selectlanguage{american}%
 & \selectlanguage{english}%
$\left|H-1\rightarrow L+1\right\rangle $ $(0.3586)$\selectlanguage{american}%
\tabularnewline
\hline 
\end{tabular}\selectlanguage{english}%
\end{table}
\bibliography{rgm}